\documentclass[11pt,aps,superscriptaddress,prc,preprint,nofootinbib]{revtex4-2}
\pdfoutput=1
\usepackage{graphicx}
\usepackage{dcolumn}
\usepackage{bm}

\usepackage{verbatim}
\usepackage{empheq}

\DeclareMathOperator{\sgn}{sgn}

\usepackage{amsmath,amsfonts,amsthm,amssymb}
\usepackage{graphics}
\usepackage{cancel}
\usepackage{multirow}
\usepackage{longtable}
\usepackage{color}
\usepackage{xcolor}
\usepackage[normalem]{ulem}
\usepackage{dsfont}

\newcounter{rownum} 
\newcommand{\Rownum}{\stepcounter{rownum} 
\arabic{rownum}.\,\,} 

\usepackage{amssymb}

\usepackage{graphicx}
\usepackage{mathtools}
\usepackage{subfigure}
\usepackage{bbold}
\usepackage{yfonts}
\usepackage[colorlinks=true,linktocpage=true,linkcolor=blue,citecolor=blue]{hyperref}
\usepackage{placeins}
\usepackage{bm} 
\usepackage{nicefrac}
\usepackage{slashed}
\usepackage{marginnote}

\newcommand{\ab}[1]{\left\langle#1\right\rangle}
\newcommand{\tr}{{\rm Tr}}

\newcommand{\EQ}[1]{Eq.~(\ref{#1})}
\newcommand{\EQn}[1]{(\ref{#1})}

\newcommand{\EQSTWO}[2]{Eqs.~(\ref{#1})~and~(\ref{#2})}

\newcommand{\EQSM}[2]{Eqs.~(\ref{#1})--(\ref{#2})}

\newcommand{\CIT}[1]{Ref.~\citep{#1}} 
\newcommand{\CITn}[1]{\citep{#1}} 

\newcommand{\av}{{\boldsymbol a}} 
\newcommand{\bv}{{\boldsymbol b}} 
 
\newcommand{\kv}{{\boldsymbol k}}

\newcommand{\wv}{{\boldsymbol w}}

\newcommand{\pv}{{\boldsymbol p}}

\newcommand{\qv}{{\boldsymbol q}}

\newcommand{\xv}{{\boldsymbol x}}

\newcommand\alphav{{\boldsymbol \alpha}}
\newcommand\gammav{{\boldsymbol \gamma}}

\newcommand\tauv{{\boldsymbol \tau}}
\newcommand\zetav{{\boldsymbol \zeta}}

\begin{document}
	
\title{Exact Wigner function for chiral spirals}

\author{Samapan Bhadury}
\email{samapan.bhadury@uj.edu.pl}
\affiliation{Institute of Theoretical Physics, Jagiellonian University, ul. St. \L ojasiewicza 11, 30-348 Kraków, Poland}

\author{Wojciech Florkowski}
\email{wojciech.florkowski@uj.edu.pl}
\affiliation{Institute of Theoretical Physics, Jagiellonian University, ul. St. \L ojasiewicza 11, 30-348 Kraków, Poland}

\author{Sudip Kumar Kar}
\email{sudip.kar@doctoral.uj.edu.pl}
\affiliation{Institute of Theoretical Physics, Jagiellonian University, ul. St. \L ojasiewicza 11, 30-348 Kraków, Poland}

\author{Valeriya Mykhaylova}
\email{valeriya.mykhaylova@uj.edu.pl}
\affiliation{Institute of Theoretical Physics, Jagiellonian University, ul. St. \L ojasiewicza 11, 30-348 Kraków, Poland}

\date{\today}

\begin{abstract}
The exact solution of the Dirac equation for fermions coupled to an external periodic chiral condensate (chiral spiral) is used to obtain the exact formula for the Wigner function (up to the quantum loop corrections). We find that the resulting expressions for various coefficients of the Wigner function exhibit properties that cannot be reproduced within the standard semiclassical expansion. The formula for the axial vector component of the Wigner function can be conveniently used to study spin polarization effects and illustrate connections between the spin density matrix and axial current. In particular, we find that during an adiabatic change of the periodic potential into a uniform one, the polarization vector is twisted from its original direction.
\end{abstract}

\maketitle
\bigskip
\section{Introduction}

In this work, we consider Dirac particles that interact with a periodic chiral condensate. The latter is described by the mean scalar and pseudoscalar fields, which in the system's rest frame vary in space as $M \cos(\qv \cdot\xv /\hbar)$ and $M \sin(\qv \cdot \xv/\hbar)$, respectively (here $\kappa=|\qv|/\hbar$ is the wave vector characterizing the space oscillation of the condensate, while $M$ is a constant). The mean field configuration of this form is known in the literature as a  chiral spiral~\CITn{Schon:2000qy,Fukushima:2010bq}. Due to the relation of the considered system to chiral quark models, we often use the name ``quarks" for the Dirac particles, although our results hold for any Dirac field coupled to the mean fields defined above. 

The interaction considered here has been studied intensively in the past~\CITn{DAUTRY1979323, Elze:1986hq, Elze:1986qd, Vasak:1987um, Elze:1987ii, BANERJEE1981326, KUTSCHERA1990566, Broniowski:1990dy, Sadzikowski:2000ap, Schon:2000qy, Sadzikowski:2006jq, Maedan:2009yi, Fukushima:2010bq, Partyka:2010em, Heinz:2013hza, Buballa:2014tba, Adhikari:2017ydi, Lakaschus:2020caq, Papadopoulos:2024agt}. Arguments have been presented that the inhomogeneous condensate lowers the energy of the ground state, compared to the homogeneous case~\CITn{DAUTRY1979323, BANERJEE1981326}. Moreover, inhomogeneous systems become naturally polarized~\CITn{DAUTRY1979323, KUTSCHERA1990566, Wang:2021owk}, and this effect may be responsible for the magnetic fields of the pulsars. All of these studies supported the idea of the formation of the pion condensate, although some recent studies disfavor the presence of chiral spirals in neutron stars~\CITn{Papadopoulos:2024agt}.

Compared to previous analyses, which extensively discussed various physical processes, our present study deliberately focuses on a specific problem. The system of quarks interacting with a periodic condensate can be treated exactly (without the quantum loop corrections). Therefore, it may be studied to demonstrate various properties of the quark Wigner function. To our knowledge, only the case of the free Dirac field has been studied at the same level of accuracy~\CITn{PhysRevC.94.024904}. The standard way to calculate the Wigner function for nontrivial systems is the semiclassical expansion (a combined expansion in the Planck constant $\hbar$ and gradients)~\cite{Florkowski:1997wu, FLORKOWSKI1996445, Weickgenannt:2019dks}. We show that the semiclassical expansion does not reproduce the properties of the exact Wigner function. This finding indicates that semiclassical analyses of complex systems must be taken with a grain of salt. 

Contrary to previous works using quark propagators in an external periodic field, our approach uses explicit expressions for the Dirac spinors. This method turns out to be very convenient for dealing with spin degrees of freedom. Our results elucidate the relations between the spin density matrix, the axial current, and the spin tensor. These outcomes may be useful for the interpretation of the spin phenomena studied recently in the context of heavy-ion collisions \cite{Liang:2004ph, Liang:2004xn, STAR:2017ckg, STAR:2018gyt, ALICE:2019aid, STAR:2019erd}. We have found interesting phenomena in the case where the periodic potential flattens out, i.e., in the limit $|\qv| \to 0$. If the system is initially characterized by some additional spin potential $\mu_{\mathfrak{s}}$, then for $|\qv| \to 0$, the spin polarization axis is twisted compared to its original direction.

The article is organized as follows: Sec. \ref{sec:Dirac_CS} introduces the Dirac equation with inhomogeneous chiral condensate and an exact solution is obtained for the spinor fields. Then, the quantization scheme of the fields is described, and the spin properties of the spinors are identified. In particular, we discuss two special cases: of the quarks being at rest and in the limit $|\qv| \to 0$. In Sec. \ref{sec:WF}, we compute the expression of the exact Wigner function and its components. The breakdown of the semiclassical expansion of the Wigner function is explained in Sec. \ref{sec:WF-sc}. We close the discussion with the conclusions and outlook in Sec. \ref{sec:C&O}. The two appendixes include the details of our calculations.

Throughout the paper, we use units where the speed of light is set equal to one, $c=1$, however, we explicitly display the Planck constant $\hbar$. In this way, we can confront our results with those obtained with the semiclassical expansion of the Wigner function. For the Levi-Civita tensor $\varepsilon^{\mu\nu\alpha\beta} $ we follow the convention $\varepsilon^{0123} = - \varepsilon_{0123} = +1$. The metric tensor is of the form $g_{\mu\nu} = \textrm{diag}(+1,-1,-1,-1)$. The scalar products for both 3- and 4-vectors are denoted by a dot, i.e., $a \cdot b = a^0 b^0 - \av \cdot \bv$.

\section{Dirac field coupled to a periodic chiral condensate}
\label{sec:Dirac_CS}

\subsection{Dirac equation and its eigenvalues}

The starting point for our investigations is the Dirac equation for the (constituent) quark field $\psi(x)$~\footnote{Note our remarks given in Introduction that the name quark is used here as a synonym of any spin~$\nicefrac{1}{2}$ fermion.}
\begin{align}
\Big[i \hbar\, \gamma_\mu\partial^\mu - \sigma - i\gamma_5 \pi \Big] \psi(x) = 0, \label{eq:Dirac-Eq}
\end{align}
where the scalar and pseudoscalar fields $\sigma$ and $\pi$ are given externally and have the form
\begin{align}
\sigma &= M \cos\left(\frac{\qv \cdot \xv }{\hbar}\right), \label{sigma-ansatz}\\
\pi &= M \sin\left(\frac{\qv \cdot \xv }{\hbar}\right). \label{pi-ansatz}
\end{align}
Here, the 3-vector $\qv$ may be treated as the spatial part of a 4-vector whose time component vanishes in the reference frame used for calculations, while the parameter $M$ is a constant, $M = \sqrt{\sigma^2 + \pi^2}$. Without loss of generality, due to rotational invariance, we take $\qv$ to be aligned along the $z$ axis and write $\qv = (0,0,q)$ with $q>0$.

For the gamma matrices, we use the Dirac representation~\cite{itzykson2006quantum},
\begin{align}
    \gamma_{0}={\begin{pmatrix}I_{2}&0\\0&-I_{2}\end{pmatrix}},\quad \gammav ={\begin{pmatrix}0&\tauv \\-\tauv &0\end{pmatrix}},\quad \gamma_{5}={\begin{pmatrix}0&I_{2}\\I_{2}&0\end{pmatrix}}, \label{gamma-matrices}
\end{align}
where $\tauv = (\tau^1, \tau^2, \tau^3)$ are the Pauli matrices and $I_2$ is the $2 \times 2$ identity matrix. Using the property of the matrix $\gamma_5$, we can combine the scalar and pseudoscalar terms into a single exponential function, which leads us to the equation
\begin{align}
    \left[i \hbar\, \gamma_\mu \partial^\mu - M e^{i \gamma_5 \left(\bm{q}\cdot \bm{x}\right)/\hbar}\right] \psi(x) &= 0. \label{Dirac-Eq2}
\end{align}
It was already found in the 1970s that this equation has analytic solutions~\cite{DAUTRY1979323}. This finding paved the way for extended studies of systems with nonuniform chiral condensates~\cite{BANERJEE1981326,KUTSCHERA1990566,Broniowski:1990dy}. In the following, we outline the construction of the solutions with ``positive'' and ``negative'' energies.


\begin{figure}[t]
    \begin{center}
        \includegraphics[scale=0.6]{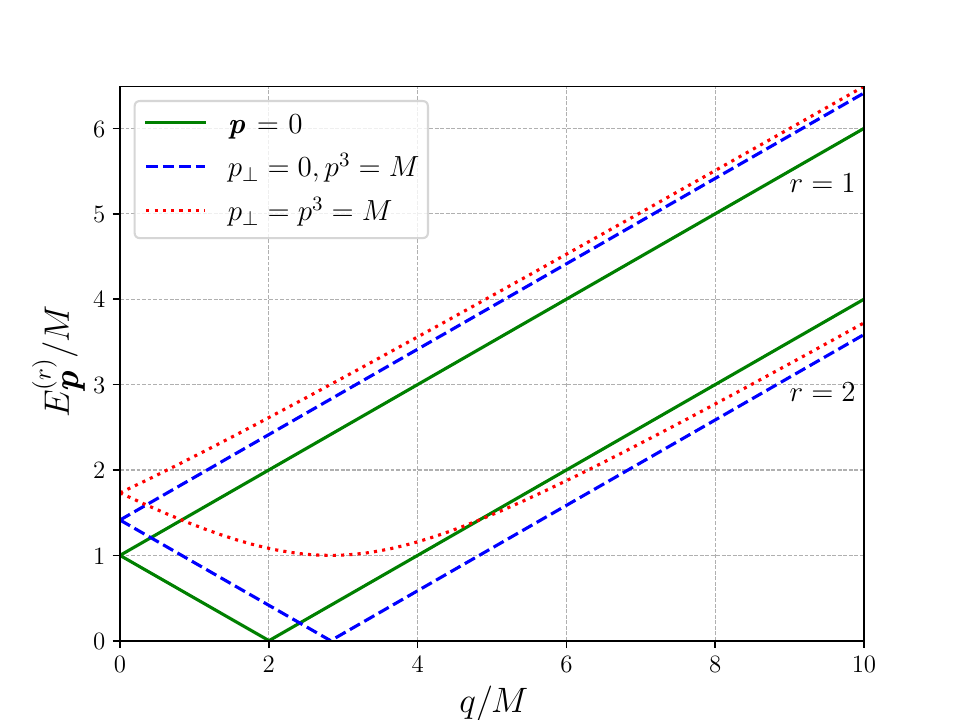}
        \caption{ \small Energy-to-mass ratio, $E_{\bm{p}}^{(r)}/M$, plotted as a function of $q/M$ for two different values of spin orientation, $r=1$ (upper lines) and $r=2$ (lower lines). The solid lines correspond to the case of particles at rest, i.e., with $\bm{p} = (0,0,0)$. The dashed lines correspond to the case of particles moving longitudinally with constant momentum equal to the effective mass, $p_\perp = 0$ and $p^3 = M$. The dotted lines correspond to the case of particles with constant 3-momentum defined by the conditions $p_\perp = M$ and $p^3 = M$.}
        \label{figure-Epr}
    \end{center}
\end{figure}

The positive-energy solutions describing particles are obtained from the ansatz,
\begin{align}
\psi_+\left(x\right) = \exp\left(-\frac{i\gamma_5}{2} \frac{\bm{q}\cdot\bm{x}}{\hbar}\right) \chi_+\left(\pv \right) e^{-ip\cdot x/\hbar},
\label{eq:posene}
\end{align}
where we assume that $p^0>0$. Substituting \EQn{eq:posene} into \EQn{eq:Dirac-Eq}, we find
\begin{align}
    \left[
    -\frac{1}{2}\boldsymbol{\alpha} \cdot\bm{q}\gamma_5+\boldsymbol{\alpha} \cdot\bm{p}+\beta M\right] \chi_+\left(\bm{p}\right) = E_\pv \, \chi_+\left(\bm{p}\right), \label{DE-chi+}
\end{align}
where we have introduced the traditional notation $\alphav=\gamma_0\gammav$ and $\beta=\gamma_0$. Since the space-time dependence disappears in the equation above, the search for the allowed energies is equivalent to finding the eigenvalues of the matrix
\begin{align}
{\begin{pmatrix} M-\frac{1}{2}\tauv \cdot \qv & \tauv \cdot \pv \\ \tauv \cdot \pv&- M-\frac{1}{2} \tauv \cdot \qv \end{pmatrix}}, \label{matrix_H+}
\end{align}
which yields two positive energies~\footnote{We note that $q$ always denotes the third component of the 3-vector $\qv$, and $q^2$ cannot be interpreted as the second component of the vector $\qv$.} \cite{DAUTRY1979323},
\begin{align}
E_\pv^{(r)} &= \sqrt{\bm{p}^2 + q^2/4 + M^2 + \left(-1\right)^{r-1} q \sqrt{M^2+ (p^3)^2}} \qquad (r=1,2). \label{E_p^r+}
\end{align}
With the help of the notation 
\begin{equation}
E_\pv^\parallel=\sqrt{M^2+(p^3)^2}\,, \quad p_\perp = \sqrt{(p^1)^2 + (p^2)^2},
\end{equation}
we may write
\begin{align}
E_\pv^{(r)} &= \sqrt{(E_\pv^\parallel+(-1)^{r-1} q/2)^2+ p_\perp^2}, \label{E_p^r++}
\end{align}
which explicitly demonstrates the positivity of energies. An interesting feature of the considered field configurations is that the energies of fermions with different spin orientations (defined by the label $r$) are different, $E_\pv^{(1)} \geq E_\pv^{(2)}$, see Fig.~\ref{figure-Epr}. This feature naturally leads to polarization phenomena that were previously studied in the context of neutron stars.


For negative-energy solutions we use the ansatz
\begin{align}
\psi_-\left(x\right) = \exp\left(-\frac{i\gamma_5}{2} \frac{\bm{q}\cdot\bm{x}}{\hbar}\right) \chi_-\left(\pv \right) e^{i p\cdot x/\hbar}, \label{eq:negene}
\end{align}
again with $p^0 > 0$. This leads to the equation
\begin{align}
    \left[\frac{1}{2}\boldsymbol{\alpha} \cdot\bm{q}\gamma_5+\boldsymbol{\alpha} \cdot\bm{p} -\beta M\right] \chi_-\left(\pv \right) &= E_\pv \, \chi_-\left(\pv \right), \label{DE-chi-}
\end{align}
and the corresponding matrix
\begin{align}
    \begin{pmatrix}
        -M+\frac{1}{2}\tauv \cdot \qv &\tauv \cdot \pv \\ \tauv \cdot \pv & M+\frac{1}{2}\tauv \cdot \qv
    \end{pmatrix}, \label{matrix_H-}
\end{align}
whose eigenvalues are the same as those given by \EQ{E_p^r+}.

\subsection{Dirac spinors}

After obtaining the eigenvalues of the Dirac equation, we solve for its eigenvectors in the two cases (for quarks and antiquarks) and obtain the following expression~\footnote{We use the notation where the normalized spinors have an additional spin index,  as compared to those used in Eqs.~\eqref{eq:posene}, \eqref{DE-chi+}, \eqref{eq:negene}, and \eqref{DE-chi-}. We also note that in the list of arguments, we do not explicitly display the dependence of the spinors on $q$, as it is externally given and fixed. },
\begin{align}
    \chi_\pm^{(r)} (\pv) =N_\pm^{(r)}\begin{pmatrix}
        \frac{E_\pv^{(r)} \pm \left(-1\right)^{r} E_\pv^{\parallel} \mp \frac{q}{2}}{p^1 + i p^2}  \\[0.5em]
        \pm \frac{p^3}{M + \left(-1\right)^{r} E_\pv^{\parallel}} \\[1em]
        \pm \frac{E_\pv^{(r)} \pm \left(-1\right)^{r} E_\pv^{\parallel} \mp \frac{q}{2}}{p^1+i p^2} \frac{p^3}{M + \left(-1\right)^{r} E_\pv^{\parallel}} \\[0.5em]
        1
    \end{pmatrix}. \label{eq:chi_pm^r}
\end{align}
Here, we have introduced the labels $\pm$ to distinguish between particles and antiparticles. The factor $N_\pm^{(r)}$ has been added to impose the normalization
\begin{align}
    {\chi_+^{\,(r)^\dagger}} (\bm{p}) \chi_+^{\,(r)} (\bm{p}) = 2 E_\pv^{(r)},
   \qquad
    {\chi_-^{\,(r)^\dagger}} (\bm{p}) \chi_-^{\,(r)} (\bm{p}) = 2 E_\pv^{(r)}. \label{chi^(r)-norm}
\end{align}
The explicit form of $ N_\pm^{(r)}$ is
\begin{align}
    N_\pm^{(r)} = \sqrt{2 E_\pv^{(r)}} \left[1 + \frac{\left(E_\pv^{(r)} \pm \left(-1\right)^r E_\pv^{\parallel} \mp \frac{q}{2}\right)^2}{(p^1)^2+(p^2)^2}\right]^{-1/2} \left[1 + \frac{(p^3)^2}{\left(M + \left(-1\right)^{r} E_\pv^{\parallel}\right)^2} \right]^{-1/2}. \label{N^(r)-def}
\end{align}


\begin{table}[t]
    \centering
    \begin{tabular}{|>{\textbf{\Rownum}}c|c|c|c|}
        \hline
        \multicolumn{1}{|c|}{} & \textbf{Chiral spiral} & \textbf{Free Dirac case}\\
        \hline
        & $\chi_{+}^{\,(r)^\dagger} (\bm{p})\, \chi_{+}^{\,(s)} (\bm{p}) = 2 E_\pv^{(r)} \delta^{rs}$ & $u^{(r)^\dagger} (\bm{p}) u^{(s)} (\bm{p}) = 2 E_\pv \delta^{rs}$\\
        \hline
        & $\chi_{-}^{\,(r)^\dagger} (\bm{p})\, \chi_{-}^{\,(s)} (\bm{p}) = 2 E_\pv^{(r)} \delta^{rs}$ & $v^{(r)^\dagger} (\bm{p}) v^{(s)} (\bm{p}) = 2 E_\pv \delta^{rs}$\\
        \hline
        & $\chi_{+}^{\,(r)^\dagger} (\bm{p})\, \chi_{-}^{\,(s)} (-\bm{p}) = 0$ & $u^{(r)^\dagger} (\bm{p}) v^{(s)} (-\bm{p}) = 0$\\
        \hline
        & $\chi_{-}^{\,(r)^\dagger} (-\bm{p})\, \chi_{+}^{\,(s)} (\bm{p}) = 0$ & $v^{(r)^\dagger} (-\bm{p}) u^{(s)} (\bm{p}) = 0$\\
        \hline
        & $\Bar{\chi}_{+}^{\,(r)} (\bm{p})\, \chi_{-}^{\,(s)} (\bm{p}) = 0$ & $\Bar{u}^{(r)} (\bm{p}) v^{(s)} (\bm{p}) = 0$ \\
        \hline
        & $\Bar{\chi}_{-}^{\,(r)} (\bm{p})\, \chi_{+}^{\,(s)} (\bm{p}) = 0$ & $\Bar{v}^{(r)} (\bm{p}) u^{(s)} (\bm{p}) = 0$ \\
        \hline
    \end{tabular}
    \caption{Orthogonality properties of the spinors used in this work, compared to the free Dirac case.}
    \label{tab:Ortho}
\end{table}

\bigskip
Spinors given by \EQ{eq:chi_pm^r} satisfy the orthogonality properties shown in the second column of Table~\ref{tab:Ortho}. Their counterparts in the free Dirac case are given in the third column. Furthermore, we derive the completeness relation (a spin sum) 
\begin{align}
    \sum_{r=1}^2 \frac{1}{2 E_\pv^{(r)}}\left[\chi_{+,\,a}^{(r)} (\bm{p}) \chi_{+,\,b}^{(r)^\dagger} (\bm{p}) + \chi_{-,\,a}^{(r)} (-\bm{p}) \chi_{-,\,b}^{(r)^\dagger} (-\bm{p}) \right] = \delta_{ab}. \label{eqn:spin_sums}
\end{align}
Here and in what follows, the small Latin letters from the beginning of the alphabet denote spinor indices.

\subsection{Second quantization}

So far, our description has used the classical field equations. To introduce a quantized field, we utilize the standard method of second quantization and use the expansion
\begin{align}
    \psi(x) = \sum_{r=1,2}\int \frac{d^3p}{(2\pi \hbar)^{3/2}} \frac{1}{\sqrt{2E_\pv^{(r)}}} \Bigl[
    u^{(r)}(\pv,\xv) b_r(\pv) e^{- \frac{i}{\hbar} p \cdot x} + v^{(r)}(\pv,\xv) c_r^\dagger(\pv) e^{\frac{i}{\hbar} p\cdot x}\Bigr],~~~~\label{psi-decomp}
\end{align}
where $b_r(\pv)$ and $c_r(\pv)$ ($b_r^\dagger(\pv)$ and $c_r^\dagger(\pv)$) are annihilation (creation) operators for particles and antiparticles. To keep the notation similar to the free Dirac case, we also define the spinors
\begin{align}
        u^{(r)}(\pv,\xv) &= \exp\left(-\frac{i \gamma_5}{2} \frac{\qv \cdot \xv}{\hbar}\right) \chi^{(r)}_+(\pv), \label{u^r}\\
        v^{(r)}(\pv,\xv) &= \exp\left(-\frac{i \gamma_5}{2} \frac{\qv \cdot \xv }{\hbar}\right) \chi^{(r)}_-(\pv). \label{v^r}
\end{align}
The field operator defined above satisfies the canonical equal-time (anti)commutation relations,
\begin{align}
    \{\psi_a(t,\bm{x}), \psi^\dagger_b(t,\bm{y})\} = \delta_{ab}\delta^{(3)}(\bm{x}-\bm{y}), \label{ETC1-field}\\
    \{\psi_a(t,\bm{x}), \psi_b(t,\bm{y})\} = \{\psi_a^\dagger(t,\bm{x}), \psi_b^\dagger(t,\bm{y})\} = 0, \label{ETC2-field}
\end{align}
provided that the creation and annihilation operators satisfy the (anti)commutation relations
\begin{align}
    &\left\{b_r(\pv), b_s^\dagger(\pv')\right\} = \left\{c_r(\pv), c_s^\dagger(\pv')\right\} = \delta^{(3)} \left(\bm{p} - \bm{p}'\right)\delta_{rs}, \label{ETC1-op}\\
    \left\{b_r(\pv), b_s(\pv')\right\} &= \left\{b_r^\dagger(\pv), b_s^\dagger(\pv')\right\} = \left\{c_r(\pv), c_s(\pv')\right\} = \left\{c_r^\dagger(\pv), c_s^\dagger(\pv')\right\} = 0. \label{ETC2-op}
\end{align}
The details of the calculations are presented in Appendix \ref{sec:ACrelnDeriv}. According to the definitions given above, the dimensions of the field operators and creation/annihilation operators are fm$^{-3/2}$ and GeV$^{-3/2}$, respectively. This is consistent with \EQ{psi-decomp}, since the dimension of the bispinors $u$ and $v$ is GeV$^{1/2}$.

 \subsection{The chiral spinors for particles at rest}

It is interesting to consider the limit of particles at rest, i.e., with a vanishing 3-momentum $\pv$. This sheds light on the connection between the index $r$ and the direction of spin polarization. As the limit $\pv \to 0$ depends on the direction in three-dimensional momentum space, we introduce spherical coordinates and write $\pv=(p^1,p^2,p^3)=(|\pv|\sin\theta \cos\phi,|\pv|\sin\theta\sin\phi,|\pv|\cos\theta)$, where $\theta \in [0,\pi]$ and $\phi \in [0,2\pi]$ are the polar and azimuthal angles, respectively. Taking the limit $|\pv| \to 0$ in our expressions for the spinors $\chi^{(r)}_\pm(\pv)$ we find
\begin{align}
\chi^{(1)}_+(\bm{p}\to 0)= - \sgn(\cos \theta) \sqrt{2 M+q} \left( \begin{array}{c} 0\\
1 \\
0\\
0
\end{array} \right), \label{chi+^(1)_p0}
\end{align}
\begin{align}
\chi^{(2)}_+(\bm{p}\to 0)= e^{-i \phi} \sqrt{| 2 M-q |} \left( \begin{array}{c} 1\\
0\\
0\\
0
\end{array} \right), \label{chi+^(2)_p0}
\end{align}

\begin{align}
\chi^{(1)}_-(\bm{p}\to 0)=e^{-i \phi} \sgn(\cos \theta) \sqrt{2 M+q}\left( \begin{array}{c} 0\\
0\\
1 \\
0
\end{array} \right), \label{chi-^(1)_p0}
\end{align}
\begin{align}
\chi^{(2)}_-(\bm{p}\to 0)= \sqrt{|2 M-q|}\left( \begin{array}{c} 0\\
0\\
0\\
1
\end{array} \right). \label{chi-^(2)_p0}
\end{align}
We observe that the spinors $\chi^{(1)}_+(\bm{p}\to 0)$ and $\chi^{(2)}_+(\bm{p}\to 0)$ describe fermions with spin down and up along the $z$ axis, respectively. Since for antiparticles the physical spin component should be considered with the opposite sign, also the spinors $\chi^{(1)}_-(\bm{p}\to 0)$ and $\chi^{(2)}_-(\bm{p}\to 0)$ describe fermions with spin down and up along the $z$-axis. We note that the overall phases appearing in the four equations above can be changed by redefinition of the original spinors. 

\subsection{The limit of a uniform condensate}

Another interesting limit to examine is the case of $q\to0$, i.e., an approach towards a uniform condensate. This limit transforms Eq.~\eqref{eq:Dirac-Eq} into the free Dirac equation for massive fermions with mass $M$. Hence, one expects that the spinors corresponding to the chiral spiral will be reduced to standard Dirac spinors in this case. Applying the limit $q\to 0$ in Eq.~\eqref{eq:chi_pm^r} and defining $E_\pv=\sqrt{\pv^2+M^2}$, we find the following expressions:
\begin{gather}
    \lim_{q\to0}~ \chi_+^{(1)}(\bm{p})= \frac{p_\perp}{\sqrt{ 2E_{\bm{p}}^{\parallel}}} 
    \begin{pmatrix}
        \frac{\sqrt{\left(E_\pv - E_{\bm{p}}^{\parallel}\right)\left(E_{\bm{p}}^{\parallel} - M\right)}}{p^1+ip^2 }\\[1.5em]
        -\frac{p^3}{\sqrt{\left(E_\pv - E_{\bm{p}}^{\parallel}\right)\left(E_{\bm{p}}^{\parallel} - M\right)}}\\[1.5em]
        -\frac{p^3\sqrt{E_\pv-E_{\bm{p}}^{\parallel}}}{(p^1+ip^2)\sqrt{E_{\bm{p}}^{\parallel} - M}}\\[1.5em]
        \frac{\sqrt{E_{\bm{p}}^{\parallel} - M}}{\sqrt{E_\pv-E_{\bm{p}}^{\parallel}}}, \label{chi+^(1)_q0}
    \end{pmatrix} ,
\end{gather}
\begin{gather}
    \lim_{q\to0}~ \chi_+^{(2)}(\bm{p})=  \frac{p_\perp}{\sqrt{ 2E_{\bm{p}}^{\parallel}}}
    \begin{pmatrix}
        \frac{\sqrt{\left(E_\pv+E_{\bm{p}}^{\parallel}\right)\left(E_{\bm{p}}^{\parallel}+M\right)}}{p^1+ip^2}\\[1.5em]
       \frac{p^3}{\sqrt{\left(E_\pv + E_{\bm{p}}^{\parallel}\right)\left(E_{\bm{p}}^{\parallel} + M\right)}}\\[1.5em]
       \frac{p^3\sqrt{E_\pv + E_{\bm{p}}^{\parallel}}}{(p^1+ip^2)\sqrt{E_{\bm{p}}^{\parallel} + M}}\\[1.5em]
        \frac{\sqrt{E_{\bm{p}}^{\parallel} + M}}{\sqrt{E_\pv + E_{\bm{p}}^{\parallel}}}, \label{chi+^(2)_q0}
    \end{pmatrix} ,
\end{gather} 
\begin{gather}
    \lim_{q\to0}~ \chi_-^{(1)}(\bm{p})=   \frac{p_\perp}{\sqrt{2E_{\bm{p}}^{\parallel}}}
    \begin{pmatrix}
        \frac{\sqrt{\left(E_\pv+E_{\bm{p}}^{\parallel}\right)\left(E_{\bm{p}}^{\parallel}-M\right)}}{p^1+ip^2}\\[1.5em]
        \frac{p^3}{\sqrt{\left(E_\pv+E_{\bm{p}}^{\parallel}\right)\left(E_{\bm{p}}^{\parallel}-M\right)}}\\[1.5em]
        \frac{p^3\sqrt{E_\pv+E_{\bm{p}}^{\parallel}}}{(p^1+ip^2)\sqrt{E_{\bm{p}}^{\parallel}-M}}\\[1.5em]
        \frac{\sqrt{E_{\bm{p}}^{\parallel}-M}}{\sqrt{E_\pv+E_{\bm{p}}^{\parallel}}}, \label{chi-^(1)_q0}
    \end{pmatrix} ,
\end{gather}
\begin{gather}
    \lim_{q\to0}~ \chi_-^{(2)}(\bm{p})=   \frac{p_\perp}{\sqrt{ 2E_{\bm{p}}^{\parallel}}}  
    \begin{pmatrix}
        \frac{\sqrt{\left(E_\pv-E_{\bm{p}}^{\parallel}\right)\left(E_{\bm{p}}^{\parallel}+M\right)}}{p^1+ip^2}\\[1.5em]
   -    \frac{p^3}{\sqrt{\left(E_\pv-E_{\bm{p}}^{\parallel}\right)\left(E_{\bm{p}}^{\parallel}+M\right)}}\\[1.5em]
       -\frac{p^3\sqrt{E_\pv-E_{\bm{p}}^{\parallel}}}{(p^1+ip^2)\sqrt{E_{\bm{p}}^{\parallel}+M}}\\[1.5em]
        \frac{\sqrt{E_{\bm{p}}^{\parallel}+M}}{\sqrt{E_\pv-E_{\bm{p}}^{\parallel}}}, \label{chi-^(2)_q0}
    \end{pmatrix} .
\end{gather}
One can easily notice that the above expressions differ from the standard formulas for the free massive Dirac spinors~\cite{itzykson2006quantum}. It turns out, however, that Eqs.~\eqref{chi+^(1)_q0}--\eqref{chi-^(2)_q0} are linear combinations of the standard spinors, see Appendix~\ref{ssec:q=0}. This behavior is similar to the effects known from the perturbation theory with degenerate levels. For our system, it implies a twist of the original polarization direction. We discuss this phenomenon in more detail below.

\section{Wigner function}
\label{sec:WF}

\subsection{Definition}

We define the Wigner function in the standard way as the Fourier transform of the expectation value of the product of two field operators taken at two different spacetime points,
\begin{align}
&W_{ab} (x,k) = \int\frac{d^4y}{(2\pi \hbar)^4} e^{-\frac{ik\cdot y}{\hbar}} \ab{\bar{\psi}_b \left(x+\frac{y}{2}\right) \psi_a\left(x-\frac{y}{2}\right)} .
\label{WF-def}
\end{align}
The above definition directly implies that
\begin{align}
& \int d^4k \, W_{ab} (x,k) = \ab{\bar{\psi}_b \left(x\right) \psi_a\left(x\right)} .
\label{WF-def-inv}
\end{align}
Using the representation of the field operators in terms of the creation and annihilation operators, we obtain
\begin{align}
&W_{ab} (x,k) = \sum_{r,s=1}^2 \!\int\!\! \frac{d^3 \bm{p}\, d^3 \bm{p}'}{\left(2\pi\hbar\right)^{3}} \!\!\int\!\! \frac{d^4y}{(2\pi \hbar)^4} \frac{e^{- \frac{i k\cdot y}{\hbar}}}{2 \sqrt{E_\pv^{(r)} E_{\bm{p}'}^{(s)}}}  \nonumber \\
\times & \left[ \bar{u}^{(s)}_b \!\left(\bm{p}'\!,\bm{x}\!+\!\frac{\bm{y}}{2}\right)\! u^{(r)}_a \!\left(\pv,\bm{x}\!-\!\frac{\bm{y}}{2}\right) \! \ab{b_s^\dagger(\pv') b_r(\pv) } 
e^{\frac{i}{\hbar} p_s'\cdot\left(x+y/2\right)}e^{- \frac{i}{\hbar} p_r\cdot\left(x-y/2\right)} \right. \nonumber\\[0.5em]
& + \left.\bar{v}^{(s)}_b \!\!\left(\bm{p}'\!,\bm{x}\!+\!\frac{\bm{y}}{2}\right)\! v^{(r)}_a \!\!\left(\bm{p},\bm{x}\!-\!\frac{\bm{y}}{2}\right)\!\! \ab{c_s (\bm{p}') c_r^\dagger (\bm{p})} e^{-\frac{i}{\hbar} p_s'\cdot\left(x+y/2\right)} e^{\frac{i}{\hbar} p_r\cdot\left(x-y/2\right)}\right]. 
\end{align}
Thus, the Wigner function depends on the expectation values of the operators $b_s^\dagger(\pv') b_r(\pv)$ and $c^\dagger_s (\bm{p}') c_r (\bm{p})$. In the following, we assume that these expectation values are diagonal in the momentum and spin space and write
\begin{align}
      \ab{b_s^\dagger (\bm{p}') b_r (\bm{p})}=\delta_{sr} \delta^{(3)} \!\left(\bm{p}'-\bm{p}\right)\! f (E_\pv^{(r)} \!-\! \mu_r), \label{part_no.-exp}\\
      \ab{c_s^\dagger (\bm{p}') c_r (\bm{p})}=\delta_{sr} \delta^{(3)} \left(\bm{p}'-\bm{p}\right)f (E_\pv^{(r)} \!+\! \bar{\mu}_r), \label{anti-part_no.-exp}
\end{align}
where the function $f$ may be interpreted as the distribution function in the momentum space (this function is assumed to be even in $\pv$). Having in mind the applications to thermal systems, we assume that $f$ depends on the particle energy and chemical potential. The latter is defined as $\mu_r=\mu_B+(-1)^{r-1}\mu_{\mathfrak{s}}$ for particles and $\bar{\mu}_r=\mu_B-(-1)^{r-1}\mu_{\mathfrak{s}}$ for antiparticles. The quantity $\mu_B$ is the baryon chemical potential (enters \EQSTWO{part_no.-exp}{anti-part_no.-exp} with the opposite sign), while $\mu_{\mathfrak{s}}$ is the spin chemical potential (enters \EQSTWO{part_no.-exp}{anti-part_no.-exp} with the same sign). We emphasize that the introduction of the spin chemical potential goes beyond the standard treatment of chiral spirals, where the spin polarization effects are solely due to the energy dependence on the spin direction, and the spin polarization vanishes in the limit $q \to 0$. In our present analysis, with finite spin chemical potential $\mu_{\mathfrak{s}}$, the system is also polarized if we consider the limit $q \to 0$. The description of the polarization modifications while the periodic potential flattens out will be discussed in more detail in the following (see Sec.~\ref{sec:A} and Appendix~\ref{ssec:q=0}). 

\bigskip 
For the sake of simplicity, we use the notation
\begin{align}
    f (E_\pv^{(r)} \!-\! \mu_r)=f^{(r)}_\pv, \qquad f (E_\pv^{(r)} \!+\! \bar{\mu}_r)=\bar{f}^{(r)}_\pv, \label{fp-def}
\end{align}
and rewrite the Wigner function as
\begin{align}
    &W_{ab} (x,k)= \!\!\sum_{r=1}^2 \!\int\!\! \frac{d^3 \bm{p}}{\left(2\pi\hbar\right)^{3}} \!\!\int\!\! \frac{d^3\bm{y}}{(2\pi \hbar)^3} \frac{1}{2 E_\pv^{(r)}} \,  e^{-i\left(\bm{p} - \bm{k}\right)\cdot \bm{y}/\hbar} \nonumber \\
    &
    \times \left[\bar{u}^{(r)}_b \!\left(\bm{p},\bm{x}\!+\!\frac{\bm{y}}{2}\right)\! u^{(r)}_a \!\left(\bm{p},\bm{x}\!-\!\frac{\bm{y}}{2}\right) f^{(r)}_\pv\delta\left(k^0 - p_r^{0}\right) \right. \nonumber\\[0.5em]
    &\quad\left.+\bar{v}^{(r)}_b \!\!\left(\!-\bm{p},\bm{x}\!+\!\frac{\bm{y}}{2}\right)\!  v^{(r)}_a \!\left(\!-\bm{p},\bm{x}\!-\!\frac{\bm{y}}{2}\right) \!\Big\{\!1 - \bar{f}^{(r)}_\pv\!\Big\} \delta\left(k^0 + p_r^0\right)\right].  \label{WF-expr1}
\end{align}
Here we have used the anticommutation relations to replace the order of the creation and annihilation operators for antiparticles, which leads to the appearance of $1$ in the expressions $1 - \bar{f}^{(r)}$. This $1$ can be eliminated if we introduce normal ordering in the definition of the Wigner function. However, we do not do this because the presence of 1 can be used to reproduce a nontrivial value of the chiral condensate from the gap equation \cite{Klevansky:1992qe}.

It is also useful to introduce the matrix notation,
\begin{align}
\begin{pmatrix} f^{(1)}_\pv  & 0 \\ 0 & f^{(2)}_\pv \end{pmatrix} =
 f_\pv 
\begin{pmatrix} 
1+ \delta f_\pv/(2  f_\pv)  & 0 \\ 0 & 1 - \delta f_\pv/(2  f_\pv) \end{pmatrix} =
 f_\pv (1 + \zetav_\pv \cdot \tauv ),
\end{align}
where  
\begin{align}
f_\pv = \frac{1}{2}(f^{(1)}_\pv+f^{(2)}_\pv),
\quad
\delta f_\pv = f^{(1)}_\pv - f^{(2)}_\pv,
\end{align}
and
\begin{align}
\zetav_\pv = (0,0,\zeta_\pv), 
\quad
\zeta_\pv = \frac{\delta f_\pv }{2 f_\pv}
= \frac{f^{(1)}_\pv - f^{(2)}_\pv}{f^{(1)}_\pv + f^{(2)}_\pv}.
\end{align}
Then we may write for particles 
\begin{align}
\ab{b_s^\dagger (\bm{p}') b_r (\bm{p})}=\delta^{(3)} 
\left(\pv'-\pv \right) f_\pv (1 + \zetav_\pv \cdot \tauv )_{sr},
\end{align}
and, similarly, for antiparticles (for which we use the same notation, however, with an additional bar)
\begin{align}
\ab{c_s^\dagger (\pv') c_r (\pv)}=\delta^{(3)} 
\left(\pv'-\pv \right) {\bar f}_\pv (1 + {\bar \zetav}_\pv \cdot \tauv )_{sr}.
\end{align}

\begin{table}[t]
    \centering
    \begin{tabular}{|c|c|c|c|c|c|}
         \hline
         & \textbf{Bi-spinor products} & \textbf{Values} & & \textbf{Bi-spinor products} & \textbf{Values}\\
         \hline
         \textbf{1}. & $\Bar{\chi}^{(r)}_{\pm} \!\left(\pm\bm{p}\right) \chi^{(r)}_{\pm} \!\left(\pm\bm{p}\right)$ & $\pm 2 M C_{r} \!\left(\bm{p}, q\right)$ &\textbf{12}. & $\Bar{\chi}^{(r)}_{\pm} \!\left(\pm\bm{p}\right) \gamma_5\, \chi^{(r)}_{\pm} \!\left(\pm\bm{p}\right)$ & 0 \\
         \hline
         \textbf{2}. & $\Bar{\chi}^{(r)}_{\pm} \!\left(\pm\bm{p}\right) \gamma_0\, \chi^{(r)}_{\pm} \!\left(\pm\bm{p}\right)$ & $2 E_{\bm{p}}^{(r)}$ & \textbf{13}. & $\Bar{\chi}^{(r)}_{\pm} \!\left(\pm\bm{p}\right) \gamma_0 \gamma_5\, \chi^{(r)}_{\pm} \!\left(\pm\bm{p}\right)$ & $2 \left(-1\right)^{r} E_{\bm{p}}^{(r)} p^3/E_{\bm{p}}^{\parallel}$ \\
         \hline
         \textbf{3}. & $\Bar{\chi}^{(r)}_{\pm} \!\left(\pm\bm{p}\right) \gamma^1\, \chi^{(r)}_{\pm} \!\left(\pm\bm{p}\right)$ & $\pm 2 p^1$ & \textbf{14}. & $\Bar{\chi}^{(r)}_{\pm} \!\left(\pm\bm{p}\right) \gamma^1 \gamma_5\, \chi^{(r)}_{\pm} \!\left(\pm\bm{p}\right)$ & $\pm 2 \left(-1\right)^{r} p^1\, p^3/E_{\bm{p}}^{\parallel}$ \\
         \hline
         \textbf{4}. & $\Bar{\chi}^{(r)}_{\pm} \!\left(\pm\bm{p}\right) \gamma^2\, \chi^{(r)}_{\pm} \!\left(\pm\bm{p}\right)$ & $\pm 2 p^2$ & \textbf{15}. & $\Bar{\chi}^{(r)}_{\pm} \!\left(\pm\bm{p}\right) \gamma^2 \gamma_5\, \chi^{(r)}_{\pm} \!\left(\pm\bm{p}\right)$ & $\pm 2 \left(-1\right)^{r} p^2\, p^3/E_{\bm{p}}^{\parallel}$ \\
         \hline
         \textbf{5}. & $\Bar{\chi}^{(r)}_{\pm} \!\left(\pm\bm{p}\right) \gamma^3\, \chi^{(r)}_{\pm} \!\left(\pm\bm{p}\right)$ & $\pm 2 p^3\, C_{r} \!\left(\bm{p}, q\right)$ & \textbf{16}. & $\Bar{\chi}^{(r)}_{\pm} \!\left(\pm\bm{p}\right) \gamma^3 \gamma_5\, \chi^{(r)}_{\pm} \!\left(\pm\bm{p}\right)$ & $\pm 2 \!\left(-1\right)^{r}\! E_{\bm{p}}^{\parallel} C_{r} \!\left(\bm{p}, q\right)$ \\
         \hline
         \textbf{6}. & $\Bar{\chi}^{(r)}_{\pm} \!\left(\pm\bm{p}\right) \Sigma^{01}\, \chi^{(r)}_{\pm} \!\left(\pm\bm{p}\right)$ & $\pm 2 \left(-1\right)^{r-1} M p^2/E_{\bm{p}}^{\parallel}\,$ & \textbf{17}. & $\Bar{\chi}^{(r)}_{\pm} \!\left(\pm\bm{p}\right) \Sigma^{01} \gamma_5\, \chi^{(r)}_{\pm} \!\left(\pm\bm{p}\right)$ & $0$ \\
         \hline
         \textbf{7}. & $\Bar{\chi}^{(r)}_{\pm} \!\left(\pm\bm{p}\right) \Sigma^{02}\, \chi^{(r)}_{\pm} \!\left(\pm\bm{p}\right)$ & $\pm 2 \left(-1\right)^{r} M p^1/E_{\bm{p}}^{\parallel}$ & \textbf{18}. & $\Bar{\chi}^{(r)}_{\pm} \!\left(\pm\bm{p}\right) \Sigma^{02} \gamma_5\, \chi^{(r)}_{\pm} \!\left(\pm\bm{p}\right)$ & $0$ \\
         \hline
         \textbf{8}. & $\Bar{\chi}^{(r)}_{\pm} \!\left(\pm\bm{p}\right) \Sigma^{03}\, \chi^{(r)}_{\pm} \!\left(\pm\bm{p}\right)$ & $0$ & \textbf{19}. & $\Bar{\chi}^{(r)}_{\pm} \!\left(\pm\bm{p}\right) \Sigma^{03} \gamma_5\, \chi^{(r)}_{\pm} \!\left(\pm\bm{p}\right)$ & $2 i \left(-1\right)^{r} M E_{\bm{p}}^{(r)}/E_{\bm{p}}^{\parallel}$\\
         \hline
         \textbf{9}. & $\Bar{\chi}^{(r)}_{\pm} \left(\pm\bm{p}\right) \Sigma^{12}\, \chi^{(r)}_{\pm} \left(\pm\bm{p}\right)$ & $2 \left(-1\right)^{r} M E_{\bm{p}}^{(r)}/E_{\bm{p}}^{\parallel}\,$ & \textbf{20}. & $\Bar{\chi}^{(r)}_{\pm} \!\left(\pm\bm{p}\right) \Sigma^{12} \gamma_5\, \chi^{(r)}_{\pm} \!\left(\pm\bm{p}\right)$ & $0$\\
         \hline
         \textbf{10}. & $\Bar{\chi}^{(r)}_{\pm} \!\left(\pm\bm{p}\right) \Sigma^{13}\, \chi^{(r)}_{\pm} \!\left(\pm\bm{p}\right)$ & $0$ & \textbf{21}. & $\Bar{\chi}^{(r)}_{\pm} \!\left(\pm\bm{p}\right) \Sigma^{13} \gamma_5\, \chi^{(r)}_{\pm} \!\left(\pm\bm{p}\right)$ & $\pm 2 i \left(-1\right)^{r} M p^1/E_{\bm{p}}^{\parallel}$\\
         \hline
         \textbf{11}. & $\Bar{\chi}^{(r)}_{\pm} \!\left(\pm\bm{p}\right) \Sigma^{23}\, \chi^{(r)}_{\pm} \!\left(\pm\bm{p}\right)$ & $0$ & \textbf{22}. & $\Bar{\chi}^{(r)}_{\pm} \!\left(\pm\bm{p}\right) \Sigma^{23} \gamma_5\, \chi^{(r)}_{\pm} \!\left(\pm\bm{p}\right)$ & $\pm 2 i \left(-1\right)^{r} M p^2/E_{\bm{p}}^{\parallel}$\\
         \hline
    \end{tabular}
    \caption{Various traces of the chiral-spiral spinors, with $C_r \left(\pv, q\right)$ defined by \EQ{eq:Cr}.}
\label{tab:chi(-2-)chi}
\end{table}

\subsection{Spinor decomposition}

Since the Wigner function is a $4\times4$ matrix in spinor indices, it has become popular and convenient to represent it as a combination of the matrices  $\Gamma=\{1,\gamma_5,\gamma_\mu,\gamma_\mu \gamma_5,\Sigma_{\mu\nu}\}$, where $\Sigma_{\mu\nu}=\frac{i}{2}[\gamma^\mu,\gamma^\nu]$. The decomposition is given as~\cite{itzykson2006quantum,Vasak:1987um,Florkowski_1996,Florkowski_1998}
\begin{align}
    W_{ab} = \left[\mathcal{F} + i \gamma_5 \mathcal{P} + \gamma_\mu \mathcal{V}^\mu +  \gamma_\mu \gamma_5 \mathcal{A}^\mu + \frac{1}{2} \Sigma_{\mu\nu} \mathcal{S}^{\mu\nu}\right]_{ab}. \label{WF-decomp}
\end{align}
Each component in this decomposition can be obtained by taking the trace of the product of the Wigner function and the corresponding element of the basis set $\Gamma$.

\subsubsection{Scalar and pseudoscalar components}

The scalar component $\mathcal{F}(x,k)$ is obtained by the direct trace of the Wigner function, leading to the expression 
\begin{align}
    \hspace{-0.3cm}\mathcal{F}(x,k) \!=\! \frac{M}{\left(2\pi\hbar\right)^3} \!\sum_{r=1}^2\! \frac{\cos\!\left(\qv \cdot \xv /\hbar \right)}{E_{\bm{k}}^{(r)}} \!\left[f^{(r)}_{\bm{k}} \delta\!\left(k^0 \!-\! E_{\bm{k}}^{(r)}\right) \!+\! \Big(\bar{f}^{(r)}_{\bm{k}} \!-\! 1\Big) \delta\!\left(k^0 + E_{\bm{k}}^{(r)}\right)\right]\! C_r \!\left(\bm{k}, \frac{q}{2}\right), \label{F-def}
\end{align}
where, for simplicity of notation, we have defined the function
\begin{align}
    C_r \left(\kv, q\right) = \left[1 + \frac{\left(-1\right)^{r-1} q}{ E_{\bm{k}}^\parallel}\right].
    \label{eq:Cr}
\end{align}
The pseudoscalar component $\mathcal{P}(x,k)$ is obtained by taking the trace of the product of $-i\gamma_5$ and the Wigner function,
\begin{align}
    \hspace{-0.3cm}\mathcal{P}(x,k) \!=\! \frac{-M}{\left(2\pi\hbar\right)^3} \!\sum_{r=1}^2\! \frac{\sin\!\left(\qv \cdot \xv/{\hbar}\right)}{E_{\bm{k}}^{(r)}} \!\left[f^{(r)}_{\bm{k}} \delta \!\left(k^0 \!-\! E_{\bm{k}}^{(r)}\right) \!+\! \Big(\bar{f}^{(r)}_{\bm{k}} \!-\! 1\Big) \delta\!\left(k^0 + E_{\bm{k}}^{(r)}\right)\right] C_r \!\left(\bm{k}, \frac{q}{2}\right).\label{P-def}
\end{align}
We emphasize that the only dependence on spacetime coordinates in $\mathcal{F}(x,k)$ and $\mathcal{P}(x,k)$ appears through the trigonometric functions $\cos(\qv \cdot\xv /\hbar)$ and $\sin(\qv \cdot \xv/\hbar)$, respectively. This dependence disappears in the linear combination 
\begin{equation}
{\cal M}(x,k) = \mathcal{F}(x,k) \sigma(x) - \mathcal{P}(x,k) \pi(x),
\label{eq:calM}
\end{equation}
which gives
\begin{align}
    \mathcal{M}(k) =  \!\sum_{r=1} ^2 \, \frac{M^2}{\left(2\pi\hbar\right)^3 E_{\bm{k}}^{(r)}}  \left[f^{(r)}_{\bm{k}}\! \delta\left(k^0 \!-\! E_{\bm{k}}^{(r)}\right) \!+\! \Big(\bar{f}^{(r)}_{\bm{k}} \!-\! 1\Big) \delta\!\left(k^0 + E_{\bm{k}}^{(r)}\right)\right] C_r \left(\bm{k}, \frac{q}{2}\right) \label{eq:calMexp}
\end{align}
and
\begin{align}
    \int d^4k\, \mathcal{M}(k) = - M^2 \,\sum_{r=1} ^2 \,\int  \frac{d^3k}{\left(2\pi\hbar\right)^3 E_{\bm{k}}^{(r)}}  \left(1 - f^{(r)}_{\bm{k}}\! \!-\bar{f}^{(r)}_{\bm{k}} \right) C_r \left(\bm{k}, \frac{q}{2}\right) .\label{eq:calMexpint}
\end{align}
Note that the integral involving $1$ is divergent. In the chiral models, such as the NJL model, this divergence is removed by introducing an ultraviolet cutoff \cite{Klevansky:1992qe, Florkowski:1997pi}.

\subsubsection{Vector components}

The vector components are obtained by taking the trace of the product of $\gamma^\mu$ and the Wigner function. The straightforward calculations lead to the following expressions:

\begin{align}
    &\mathcal{V}^0(k) = \sum_{r=1}^2 \!\frac{1}{\left(2\pi\hbar\right)^{3}} \bigg\{\!\left[ \delta \!\left(k^0 \!-\! 
     E_{ \kv +\frac{\qv}{2} }^{(r)}\right)\! f_{\kv +\frac{\qv}{2}}^{(r)} \!-\! \delta \!\left(k^0 \!+\! E_{\kv +\frac{\qv}{2}}^{(r)}\right) \Big(\!\bar{f}_{\kv +\frac{\qv}{2}}^{\,(r)} \!-\! 1\!\Big) \right]\! \frac{C_{r} \!\left(\kv +\frac{\qv}{2}, k^3 + \frac{q}{2}\right)}{2} \nonumber\\
    &\hspace{1.6cm} + \!\left[\delta \!\left(k^0 \!-\! E_{\kv -\frac{\qv}{2}}^{(r)}\right)\! f_{\kv -\frac{\qv}{2}}^{(r)} \!-\! \delta \!\left(k^0 \!+\! E_{\kv -\frac{\qv}{2}}^{(r)}\right) \!\Big(\!\bar{f}_{\kv -\frac{\qv}{2}}^{\,(r)} \!-\! 1 \!\Big) \right]\! \frac{C_{r-1} \!\left(\kv -\frac{\qv}{2},  k^3 - \frac{q}{2}\right)}{2} \!\!\bigg\}, \label{V0-def}
\end{align}

\begin{align}
    &\mathcal{V}^1(k) \!=\! \sum_{r=1}^2 \!\!\frac{k^1}{\left(2 \pi \hbar\right)^3} \!\!\left\{\!\left[\delta\left(k^0 \!-\! E_{\kv +\frac{\qv}{2}}^{(r)}\right)\! f_{\kv +\frac{\qv}{2}}^{(r)} \!-\! \delta \!\left(k^0 \!+\! E_{\kv +\frac{\qv}{2}}^{(r)}\right)\!\Big(\bar{f}_{\kv +\frac{\qv}{2}}^{(r)} \!-\! 1 \Big)\right] \frac{C_{r} \!\left(\kv +\frac{\qv}{2}, k^3 + \frac{q}{2}\right)}{2 k^0}\right. \nonumber\\
    &\hspace{1.6cm}+ \left.\!\left[\delta \!\left(k^0 \!-\! E_{\kv -\frac{\qv}{2}}^{(r)}\right)\! f_{\kv -\frac{\qv}{2}}^{(r)} -\! \delta \!\left(k^0 \!+\! E_{\kv -\frac{\qv}{2}}^{(r)}\right)\!\!\Big(\bar{f}_{\kv -\frac{\qv}{2}}^{(r)} \!-\! 1 \Big)\right] \frac{C_{r-1} \!\left(\kv -\frac{\qv}{2}, k^3 - \frac{q}{2}\right)}{2 k^0}\right\}, \label{V1-def}
\end{align}
\begin{align}
    &\mathcal{V}^2(k) \!=\! \sum_{r=1}^2 \!\!\frac{k^2}{\left(2 \pi \hbar\right)^3} \!\!\left\{\!\left[\delta\left(k^0 \!-\! E_{\kv +\frac{\qv}{2}}^{(r)}\right)\! f_{\kv +\frac{\qv}{2}}^{(r)} \!-\! \delta \!\left(k^0 \!+\! E_{\kv +\frac{\qv}{2}}^{(r)}\right)\!\Big(\bar{f}_{\kv +\frac{\qv}{2}}^{(r)} \!-\! 1\Big)\right] \frac{C_{r} \!\left(\kv +\frac{\qv}{2}, k^3 + \frac{q}{2}\right)}{2 k^0}\right. \nonumber\\
    &\hspace{1.6cm}+\! \left.\!\left[\delta \!\left(k^0 \!-\! E_{\kv -\frac{\qv}{2}}^{(r)}\right)\! f_{\kv -\frac{\qv}{2}}^{(r)} \!-\! \delta \!\left(k^0 \!+\! E_{\kv -\frac{\qv}{2}}^{(r)}\right)\!\!\Big(\bar{f}_{\kv -\frac{\qv}{2}}^{(r)} \!-\! 1\Big)\right] \frac{C_{r-1} \!\left(\kv -\frac{\qv}{2}, k^3 - \frac{q}{2}\right)}{2 k^0}\right\}, \label{V2-def}
\end{align}

\begin{align}
    &\mathcal{V}^3(k)= \sum_{r=1}^2 \frac{1}{\left(2\pi\hbar\right)^{3}} \Bigg\{\!\left[ \delta\left(k^0 - E_{\kv +\frac{\qv}{2}}^{(r)}\right) f_{\kv +\frac{\qv}{2}}^{(r)} - \delta\left(k^0 + E_{\kv +\frac{\qv}{2}}^{(r)}\right)\!\Big(\!\bar{f}_{\kv +\frac{\qv}{2}}^{(r)} \!-\! 1\!\Big)\right] \nonumber\\
    &\hspace{3.5cm}\times \left[k^3-\left(-1\right)^{r}E_{\kv +\frac{\qv}{2}}^{\parallel}\right]\! \frac{C_{r} \!\left(\kv +\frac{\qv}{2}, \frac{q}{2}\right)}{2k^0} \notag\\
    &\hspace{3cm}+ \left[\delta\left(k^0 - E_{\kv -\frac{\qv}{2}}^{(r)}\right) f_{\kv -\frac{\qv}{2}}^{(r)} - \delta\left(k^0 + E_{\kv -\frac{\qv}{2}}^{(r)}\right)\!\Big(\!\bar{f}_{\kv -\frac{\qv}{2}}^{(r)} \!-\! 1\!\Big)\right] \nonumber\\ 
    &\hspace{3.5cm}\times\left[k^3+\left(-1\right)^{r}E_{\kv -\frac{\qv}{2}}^{\parallel}\right]\! \frac{C_{r-1} \!\left(\kv -\frac{\qv}{2}, \frac{q}{2}\right)}{2 k^0}\!\Bigg\} .\label{V3-def}
\end{align}
Here, $E_{ \kv +\frac{\qv}{2} }^{(r)}$ is defined, in accordance with \EQn{E_p^r+}, as
\begin{align}
E_{ \kv +\frac{\qv}{2} }^{(r)} &= \sqrt{\left(\kv +\frac{\qv}{2}\right)^2 + q^2/4 + M^2 + \left(-1\right)^{r-1} q \sqrt{M^2+ \left(p^3+\frac{q}{2}\right)^2 }} \qquad (r=1,2). \label{E_pq^r+}
\end{align}
One can check that the integrals of the vector components over the momentum space vanish for the spatial components, with the zeroth component given by the integral
\begin{align}
   \mathcal{V}^0(x) = \int d^4k \, \mathcal{V}^0(x,k) = \sum_{r=1}^2  \int \frac{d^3 k}{\left(2\pi\hbar\right)^3} \left(f_{\bm{k}}^{(r)} - \bar{f}_{\bm{k}}^{(r)}\right).
    \label{eq:V0int}
\end{align}
Here, the divergent term in the integral has been ignored. As expected, \EQ{eq:V0int} defines the density $\mathcal{V}^0(x)$ as the difference between the particle and antiparticle contributions. 

\subsubsection{Axial vector components}
\label{sec:A}

The axial vector component of the Wigner function is obtained by taking the trace of the product of $\gamma_5\gamma^\mu$ and $W(x,k)$. In this way, we find
\begin{align}
    &\mathcal{A}^0= \sum_{r=1}^2 \!\frac{1}{\left(2\pi\hbar\right)^{3}} \bigg\{\!\left[ \delta \!\left(k^0 \!-\! E_{\kv+\frac{\qv}{2}}^{(r)}\right)\! f_{\kv+\frac{\qv}{2}}^{(r)} \!-\! \delta \!\left(k^0 \!+\! E_{\kv+\frac{\qv}{2}}^{(r)}\right) \Big(\!\bar{f}_{\kv+\frac{\qv}{2}}^{\,(r)} \!-\! 1\!\Big) \right]\! \frac{C_{r} \!\left(\kv + \frac{\qv}{2}, k^3 + \frac{q}{2}\right)}{2} \nonumber\\
    &\hspace{2cm}- \left[\delta \!\left(k^0 \!-\! E_{\kv-\frac{\qv}{2}}^{(r)}\right)\! f_{\kv-\frac{\qv}{2}}^{(r)} \!-\! \delta \!\left(k^0 \!+\! E_{\kv-\frac{\qv}{2}}^{(r)}\right) \!\Big(\!\bar{f}_{\kv-\frac{\qv}{2}}^{\,(r)} \!-\! 1 \!\Big) \right]\! \frac{C_{r-1} \!\left(\kv-\frac{\qv}{2}, k^3 - \frac{q}{2}\right)}{2} \!\!\bigg\} ,\label{A0-def}
\end{align}
    
\begin{align}
    &\mathcal{A}^1 \!=\! \sum_{r=1}^2 \!\!\frac{k^1}{\left(2 \pi \hbar\right)^3} \!\!\left\{\!\left[\delta\left(k^0 \!-\! E_{\kv+\frac{\qv}{2}}^{(r)}\right)\! f_{\kv+\frac{\qv}{2}}^{(r)} \!-\! \delta \!\left(k^0 \!+\! E_{\kv+\frac{\qv}{2}}^{(r)}\right)\!\Big(\bar{f}_{\kv+\frac{\qv}{2}}^{(r)} \!-\! 1 \Big)\right] \frac{C_{r} \!\left(\kv +\frac{\qv}{2}, k^3 + \frac{q}{2}\right)}{2 k^0}\right. \nonumber\\
    &\hspace{1.8cm}- \left.\!\left[\delta \!\left(k^0 \!-\! E_{\kv-\frac{\qv}{2}}^{(r)}\right)\! f_{\kv-\frac{\qv}{2}}^{(r)} -\! \delta \!\left(k^0 \!+\! E_{\kv-\frac{\qv}{2}}^{(r)}\right)\!\!\Big(\bar{f}_{\kv-\frac{\qv}{2}}^{(r)} \!-\! 1 \Big)\right] \frac{C_{r-1} \!\left(\kv - \frac{\qv}{2}, k^3 - \frac{q}{2}\right)}{2 k^0}\right\}, \label{A1-def}
\end{align}
    
\begin{align}
    &\mathcal{A}^2 \!=\! \sum_{r=1}^2 \!\!\frac{k^2}{\left(2 \pi \hbar\right)^3} \!\!\left\{\!\left[\delta\left(k^0 \!-\! E_{\kv+\frac{\qv}{2}}^{(r)}\right)\! f_{\kv+\frac{\qv}{2}}^{(r)} \!-\! \delta \!\left(k^0 \!+\! E_{\kv+\frac{\qv}{2}}^{(r)}\right)\!\Big(\bar{f}_{\kv+\frac{\qv}{2}}^{(r)} \!-\! 1\Big)\right] \frac{C_{r} \,\left(\kv+\frac{\qv}{2}, k^3 + \frac{q}{2}\right)}{2 k^0}\right. \nonumber\\
    &\hspace{1.8cm}- \left.\!\left[\delta \!\left(k^0 \!-\! E_{\kv-\frac{\qv}{2}}^{(r)}\right)\! f_{\bm{k-\frac{q}{2}}}^{(r)} -\! \delta \!\left(k^0 \!+\! E_{\kv-\frac{\qv}{2}}^{(r)}\right)\!\!\Big(\bar{f}_{\kv-\frac{\qv}{2}}^{(r)} \!-\! 1\Big)\right] \frac{C_{r-1} \!\left(\kv-\frac{\qv}{2}, k^3 - \frac{q}{2}\right)}{2 k^0}\right\}, \label{A2-def}
\end{align}
    
\begin{align}
    &\mathcal{A}^3= \sum_{r=1}^2  \frac{1}{\left(2\pi\hbar\right)^{3}} \Bigg\{\!\! \left[\delta\left(k^0 - E_{\kv+\frac{\qv}{2}}^{(r)}\right) f_{\kv+\frac{\qv}{2}}^{(r)} \!-\! \delta\left(k^0 + E_{\kv+\frac{\qv}{2}}^{(r)}\right)\!\Big(\!\bar{f}_{\kv+\frac{\qv}{2}}^{(r)} \!-\!1 \!\Big)\right] \nonumber\\
    &\hspace{3.5cm}\times \left[k^3-\left(-1\right)^{r} E_{\kv+\frac{\qv}{2}}^{\parallel} \right] \frac{C_{r} \!\left(\kv+\frac{\qv}{2}, \frac{q}{2}\right)}{2k^0} \notag\\
    &\hspace{3cm}- \left[\delta\left(k^0 - E_{\kv-\frac{\qv}{2}}^{(r)}\right) f_{\kv-\frac{\qv}{2}}^{(r)} \!-\! \delta\left(k^0 + E_{\kv-\frac{\qv}{2}}^{(r)}\right)\!\Big(\!\bar{f}_{\kv-\frac{\qv}{2}}^{(r)} \!-\!1 \!\Big)\right] \nonumber\\
    &\hspace{3.5cm}\times \left[k^3+\left(-1\right)^{r} E_{\kv-\frac{\qv}{2}}^{\parallel}\right] \frac{C_{r} \!\left(\kv-\frac{\qv}{2}, \frac{q}{2} \right)}{2k^0} \!\Bigg\}. \label{A3-def}
\end{align}

In the case of the axial current, $A^\mu=\langle\bar{\psi}\gamma^\mu\gamma^5\psi\rangle$ which may also be obtained by integrating the axial vector component of the wigner function over the 4-momentum space, the integrals for the zeroth, first, and second components. The only non-zero result is found for the third component, reading 
\begin{align}
     A^3(x) = \int d^4k \, \mathcal{A}^3(x,k) = \sum_{r=1}^2  \int \frac{d^3 k}{\left(2\pi\hbar\right)^3} \frac{\left(-1\right)^r E_{\bm{k}}^{\parallel}}{E_{\bm{k}}^{(r)}} \left[1 + \frac{\left(-1\right)^{r-1} q}{2 E_{\bm{k}}^{\parallel}}\right] \left(f^{(r)}_{\bm{k}} + \bar{f}^{(r)}_{\bm{k}}\right). \label{eqn:A3_full}
\end{align}
Here again,  the vacuum term has been ignored. The expression above agrees with the formula obtained in~\cite{KUTSCHERA1990566} up to the internal degrees of freedom connected with flavor and color.

\begin{figure}[t]
    \begin{center}
        \includegraphics[scale=0.6]{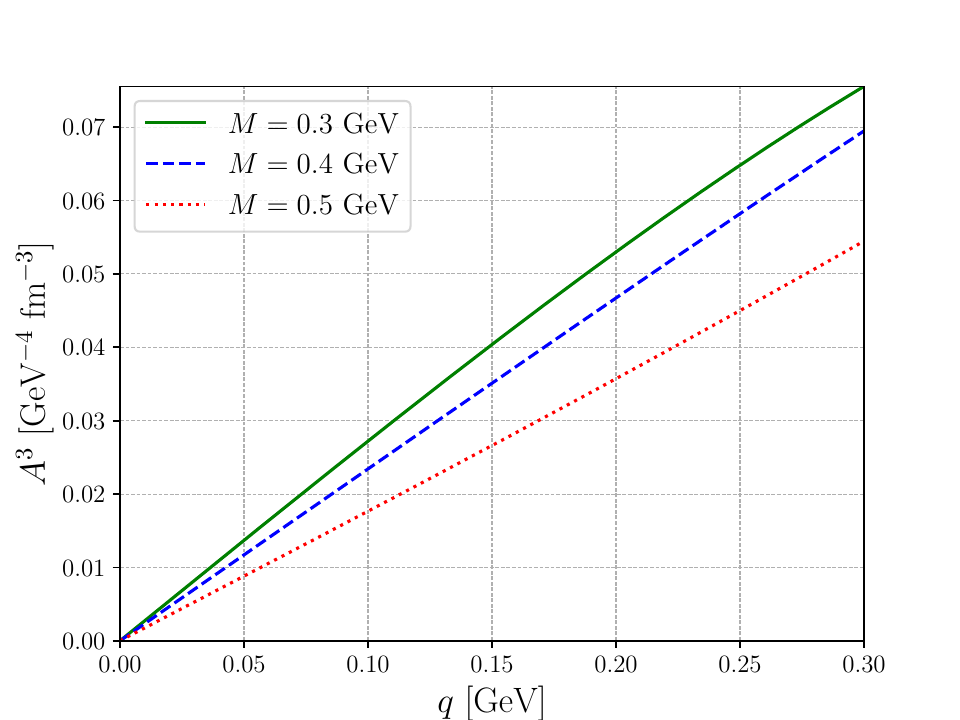}
        \caption{ \small Third component of the axial vector as a function of the inhomogeneity factor $q$ for different values of effective masses at constant chemical potential and temperature of $\mu = 0.1$~GeV and $T=0.15$~GeV, respectively. }
        \label{figure-A3_q}
    \end{center}
\end{figure}

\begin{figure}[t]
    \begin{center}
        \includegraphics[scale=0.6]{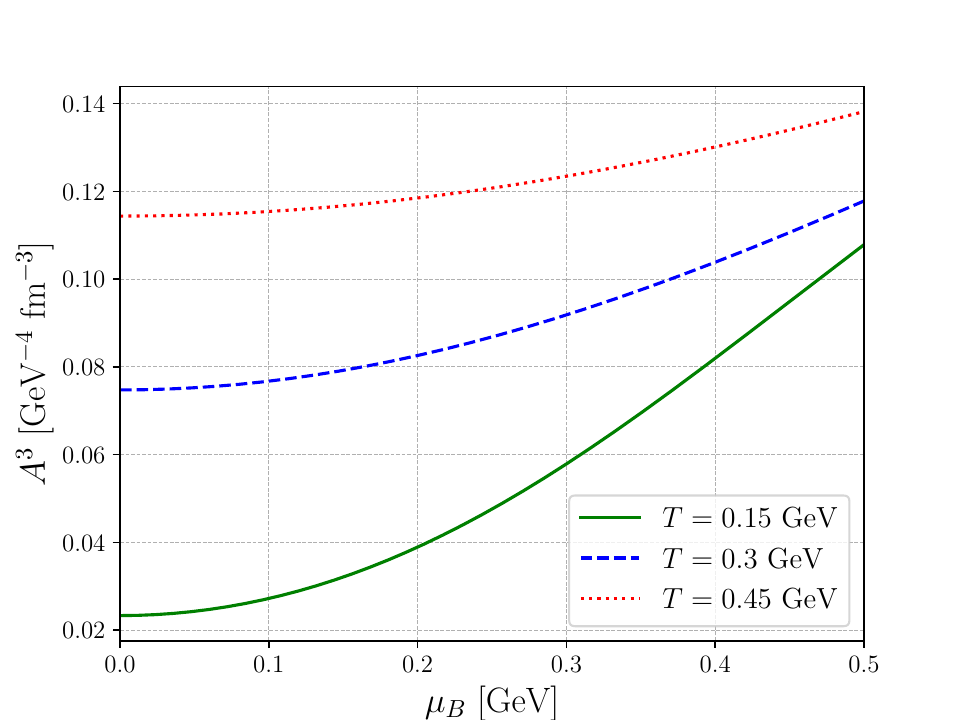}
        \caption{ \small Third component of the axial vector as a function of the chemical potential $\mu$ for different values of temperature at constant mass and inhomogeneity factor of $M = 0.3$~GeV and $q=0.1$~GeV, respectively.}
        \label{figure-A3_mu}
    \end{center}
\end{figure}

The knowledge of the axial current allows us to directly calculate the spin densities. In the canonical formulation, the spin tensor $S_{\rm can}^{\alpha \beta \gamma}$ is defined as the contraction of the Levi-Civita tensor and the axial current \cite{Florkowski:2018ahw}, $S_{\rm can}^{\alpha \beta \gamma} = -\frac{1}{2} \varepsilon^{\alpha \beta \gamma \delta} A_\delta$. This implies that the density of the third spin component is $S^3 \equiv S_{\rm can}^{0 1 2} = -\frac{1}{2} \varepsilon^{0 1 2 3} A_3 = \frac{1}{2} A^3$. The values of $A^3$ plotted as functions of the inhomogeneity factor $q$ and baryon potential $\mu_B$ at a vanishing spin potential are shown in Figs.~\ref{figure-A3_q} and~\ref{figure-A3_mu} respectively.

It is interesting to consider the limit $q \to 0$ of \EQ{eqn:A3_full}. If the spin chemical potential vanishes, $\mu_{\mathfrak{s}}=0$, the contributions from opposite spins cancel with each other, and $\mathcal{A}^3(x)=0$. However, for the nonzero spin chemical potential, the result for $\mathcal{A}^3(x)$ may be different from zero. In this case, it is also interesting to give the formula for the phase space density of the axial current:
\begin{align}
    &\mathcal{A}^0= \sum_{r=1}^2\!\frac{2}{\left(2\pi\hbar\right)^{3}}\delta(k^2-M^2)(-1)^{r+1}\Bigl(\frac{k^3k^0}{E^\parallel_{\bm{k}}}\Bigr)\Biggl[\Theta(k^0) f_{\bm{k}}^{(r)}  - \Theta(-k^0)\!\Big\{\!1 - \bar{f}_{\bm{k}}^{(r)}\Big\}\Biggr],\\
     &\mathcal{A}^1= \sum_{r=1}^2\!\frac{2}{\left(2\pi\hbar\right)^{3}}\delta(k^2-M^2)\frac{(-1)^{r+1}k^1k^3}{E_{\bm{k}}^{\parallel}}\Biggl[\Theta(k^0) f_{\bm{k}}^{(r)}  - \Theta(-k^0)\!\Big\{\!1 - \bar{f}_{\bm{k}}^{(r)}\Big\}\Biggr],\\
     &\mathcal{A}^2= \sum_{r=1}^2\!\frac{2}{\left(2\pi\hbar\right)^{3}}\delta(k^2-M^2)\frac{(-1)^{r+1}k^2k^3}{E_{\bm{k}}^{\parallel}}\Biggl[\Theta(k^0) f_{\bm{k}}^{(r)}  - \Theta(-k^0)\!\Big\{\!1 - \bar{f}_{\bm{k}}^{(r)}\Big\}\Biggr],\\
     &\mathcal{A}^3= \sum_{r=1}^2\!\frac{2}{\left(2\pi\hbar\right)^{3}}\delta(k^2-M^2)(-1)^{r+1}E_{\bm{k}}^{\parallel}\Biggl[\Theta(k^0) f_{\bm{k}}^{(r)}  - \Theta(-k^0)\!\Big\{\!1 - \bar{f}_{\bm{k}}^{(r)}\Big\}\Biggr].
\end{align}
We observe that the phase-space density of the axial current meets the condition $k_\mu \mathcal{A}^\mu~=~0$. This is so since the axial current is proportional to the 4-vector $\left(k^3 E_{\bm{k}}^{\parallel}, k^1 k^3, k^2 k^3, (E_{\bm{k}}^{\parallel})^2\right)$. The appearance of this vector can be interpreted as a twist of the original polarization axis directed along the $z$ axis, while the limit $q \to 0$ is taken. This behavior is discussed in detail in Appendix~\ref{ssec:q=0}.

\subsubsection{Tensor components}

Finally, the tensor components are computed by taking the trace of the product of $\Sigma^{\mu\nu}$ and $W(x,k)$:
\begin{align}
    \mathcal{S}^{01} &=\frac{M k^2}{E_{\bm{k}}^{\parallel}} \cos\left(\frac{\bm{q}\cdot\bm{x}}{\hbar}\right) \!\!\sum_{r=1}^2\! \frac{\left(-1\right)^{r-1}}{\left(2\pi\hbar\right)^3} \!\!\frac{1}{k^0} \left[\delta\left(k^0 \!-\! E_{\bm{k}}^{(r)}\right) f^{(r)}_{\bm{k}} \! - \! \delta\!\left(k^0 \!+\! E_{\bm{k}}^{(r)}\right) \Big(\bar{f}^{(r)}_{\bm{k}} \!-\! 1\Big)\right], \label{S01-def} \\[1em]
    \mathcal{S}^{02} &= \frac{M k^1}{E_{\bm{k}}^{\parallel}} \cos\!\left(\frac{\bm{q}\cdot\bm{x}}{\hbar}\right) \!\!\sum_{r=1}^2 \frac{\left(-1\right)^{r}}{\left(2\pi\hbar\right)^3} \frac{1}{k^0} \left[\delta\!\left(k^0 \!-\! E_{\bm{k}}^{(r)}\right) f^{(r)}_{\bm{k}} \! - \! \delta\!\left(k^0 \!+\! E_{\bm{k}}^{(r)}\right) \left(\bar{f}^{(r)}_{\bm{k}} \!-\! 1\right)\right], \label{S02-def}
\end{align}
\begin{align}
    \mathcal{S}^{03} &= \frac{M}{E_{\bm{k}}^{\parallel}} \sin\!\left(\frac{\bm{q}\cdot\bm{x}}{\hbar}\right)\!\!\sum_{r=1}^2 \frac{\left(-1\right)^{r}}{\left(2\pi\hbar\right)^3} \left[\delta\left(k^0 - E_{\bm{k}}^{(r)}\right) f^{(r)}_{\bm{k}} - \delta\left(k^0 + E_{\bm{k}}^{(r)}\right) \left(\bar{f}^{(r)}_{\bm{k}} - 1\right)\right], \label{S03-def}\\[1em]
    \mathcal{S}^{12} &= \frac{M}{E_{\bm{k}}^{\parallel}} \cos\!\left(\frac{\bm{q}\cdot\bm{x}}{\hbar}\right) \!\!\sum_{r=1}^2 \frac{\left(-1\right)^{r}}{\left(2\pi\hbar\right)^3} \left[\delta\left(k^0 - E_{\bm{k}}^{(r)}\right) f^{(r)}_{\bm{k}} - \delta\left(k^0 + E_{\bm{k}}^{(r)}\right) \left(\bar{f}^{(r)}_{\bm{k}} - 1\right)\right], \label{S12-def}\\[1em]
    \mathcal{S}^{13} &= \frac{M k^1}{E_{\bm{k}}^\parallel} \sin\!\left(\frac{\bm{q}\cdot\bm{x}}{\hbar}\right) \!\!\sum_{r=1}^2\! \frac{\left(-1\right)^{r}}{\left(2\pi\hbar\right)^3} \frac{1}{k^0} \left[\delta\left(k^0 \!-\! E_{\bm{k}}^{(r)}\right) f^{(r)}_{\bm{k}} \! - \! \delta\left(k^0 + E_{\bm{k}}^{(r)}\right) \left(\bar{f}^{(r)}_{\bm{k}} \!-\! 1\right)\right], \label{S13-def}\\[1em]
    \mathcal{S}^{23} &= \frac{M k^2}{E_{\bm{k}}^\parallel} \sin\!\left(\frac{\bm{q}\cdot\bm{x}}{\hbar}\right) \!\!\sum_{r=1}^2\! \frac{\left(-1\right)^{r}}{\left(2\pi\hbar\right)^3} \frac{1}{k^0} \left[\delta\left(k^0 \!-\! E_{\bm{k}}^{(r)}\right) f^{(r)}_{\bm{k}} \! - \! \delta\left(k^0 + E_{\bm{k}}^{(r)}\right) \left(\bar{f}^{(r)}_{\bm{k}} \!-\! 1\right)\right]. \label{S23-def}
\end{align}

\section{Semiclassical expansion of the Wigner function}%
\label{sec:WF-sc}

\subsection{General setup}%
\label{sec:WF-sc-setup}

The correlator $\ab{\bar{\psi}_b \left(y\right) \psi_a\left(x\right)} $ appearing in the definition of the Wigner function satisfies the same Dirac equation as the field $\psi_a\left(x\right)$. Therefore, one can make the Wigner transform of the Dirac equation satisfied by the correlator and obtain an equation satisfied by the Wigner function itself. In practice, this method is useful for weakly inhomogeneous systems where one can restrict oneself to first-order gradients in position and momentum space. To be more specific, such gradients are typically multiplied by the Planck constant, which introduces a dimensionless operator that includes $\hbar$, $\partial/\partial x^\mu$, and $\partial/\partial k^\mu$. This approach is known as the semiclassical expansion and has become a common tool to describe nonequilibrium dynamics \cite{Vasak:1987um}. 

In the case of the periodic chiral condensate considered in this work, the semiclassical expansion leads to the equation \cite{Florkowski_1996,Florkowski_1998}
\begin{align}
    \left[ \left( k^\mu + \frac{i\hbar}2\partial^\mu \right)\gamma_\mu - \sigma(x) + \frac{i\hbar}{2}\partial_\mu\sigma(x)\partial^\mu_k -i\gamma_5\pi(x)-\frac{\hbar}{2}\gamma_5\partial_\mu\pi(x)\partial^\mu_k\right]W(x,k)=0.
\end{align}
In the next step, it is common to use the decomposition of the Wigner function defined by \EQ{WF-decomp}, which leads to a system of coupled equations of the form:
\begin{align}
    &K^\mu \mathcal{V}_\mu - \sigma \mathcal{F} + \pi \mathcal{P} = \frac{i \hbar}{2} \Big[\left(\partial_\mu \pi\right) \left(\partial_k^\mu \mathcal{P}\right) - \left(\partial_\mu \sigma\right) \left(\partial_k^\mu \mathcal{F}\right) \Big], \label{KE-WF1}\\[1em]
    &- i K^\mu \mathcal{A}_\mu - \sigma \mathcal{P} - \pi \mathcal{F} = - \frac{i \hbar}{2} \Big[\left(\partial_\mu \pi\right) \left(\partial_k^\mu \mathcal{F}\right) + \left(\partial_\mu \sigma\right) \left(\partial_k^\mu \mathcal{P}\right) \Big], \label{KE-WF2}\\[1em]
    &K_\mu \mathcal{F} + i K^\nu \mathcal{S}_{\nu\mu} - \sigma \mathcal{V}_\mu + i \pi \mathcal{A}_\mu = \frac{i \hbar}{2} \Big[i \left(\partial_\nu \pi\right) \left(\partial_k^\nu \mathcal{A}_\mu\right) - \left(\partial_\nu \sigma\right) \left(\partial_k^\nu \mathcal{V}_\mu\right) \Big], \label{KE-WF3}\\[1em]
    &i K^\mu \mathcal{P} - K_\nu \Tilde{\mathcal{S}}^{\nu\mu} - \sigma \mathcal{A}^\mu + i \pi \mathcal{V}^\mu = \frac{i \hbar}{2} \Big[i \left(\partial_\nu \pi\right) \left(\partial_k^\nu \mathcal{V}^\mu\right) - \left(\partial_\nu \sigma\right) \left(\partial_k^\nu \mathcal{A}^\mu\right) \Big], \label{KE-WF4}\\[1em]
    &i\! \left(\!K^\mu \mathcal{V}^\nu\! - \!K^\nu \mathcal{V}^\mu\right)\! - \!\epsilon^{\mu\nu\tau\sigma} K_\tau \mathcal{A}_\sigma \!-\! \pi \Tilde{\mathcal{S}}^{\mu\nu} \!+\! \sigma \mathcal{S}^{\mu\nu} \!=\! \frac{i \hbar}{2} \Big[\left(\partial_\gamma \sigma\right) \left(\partial_k^\gamma \mathcal{S}^{\mu\nu}\right)\! - \!\left(\partial_\gamma \pi\right) \left(\partial_k^\gamma \Tilde{\mathcal{S}}^{\mu\nu}\right)\Big]. \label{KE-WF5} 
\end{align}
%
Here we have defined the operator $K^\mu = k^\mu + \frac{i \hbar}{2} \partial^\mu$ and $\tilde{\mathcal{S}}$ is the dual tensor to the tensor $\mathcal{S}$, namely
\begin{align}
    \tilde{\mathcal{S}}^{\mu\nu} = \frac12 \epsilon^{\mu\nu\alpha\beta}\mathcal{S}_{\alpha\beta}.
\end{align}
Since the system of the equations introduced above is still quite complicated, one solves it by expanding all the coefficients of the Wigner function as a series in $\hbar$, 
\begin{align}
     C=C_{(0)}+\hbar C_{(1)}+\hbar^2C_{(2)}+...\quad,
\end{align}
where $C$ is one of the coefficients from the set $\{ {\cal F}, {\cal P}, {\cal V}^\mu, {\cal A}^\mu, {\cal S}^{\mu\nu} \}$. 

\subsection{Expansion in \texorpdfstring{$\hbar$}{~}}%
\label{sec:WF-sc-hbar}

The coefficients $C_{(0)}$ are commonly interpreted as the limit of the coefficients $C$ for $\hbar~\to~0$. However, we know from our exact calculations presented in the previous sections that all the coefficients of the Wigner function include the factor $h^3 = (2\pi \hbar)^3$ in the denominator. Consequently, the formal limit $ \hbar \to 0$ cannot be taken. This issue can be resolved if the factors $h^{-3}$ are absorbed into the definitions of the distribution functions. In fact, we have defined the distribution functions $f$ as dimensionless quantities. Thus, the combinations $F = f/h^3$ become the real phase space distribution functions that define the number of particles $\Delta N$ in the element of phase space $\Delta^3x \Delta^3p$ (with the dimension GeV$^{-3}$~fm$^{-3}$). In the following, we express the Wigner functions in terms of the distributions $F$ assuming that the $F$'s have the well-defined limit as $\hbar \to 0$.~\footnote{Formally, this corresponds to taking the limit $\hbar \to 0$ with the fixed ratio $F = f/h^3$.} Then, the vector and axial vector components of the Wigner function become independent of $\hbar$, while the scalar, pseudoscalar and tensor components become of the form $g(k) \sin(\qv \cdot \xv /\hbar) $ or $g(k) \cos(\qv \cdot \xv /\hbar)$, where $g(k)$ is independent of $\hbar$.

Although the functions $\sin(\qv \cdot \xv /\hbar)$ and $\cos(\qv \cdot \xv /\hbar)$ have no well-defined expansion for small values of $\hbar$ either, we can still use \EQSM{KE-WF1}{KE-WF5} and make an expansion in $\hbar$ -- the space derivatives of the mean fields are multiplied by $\hbar$, which cancels the $1/\hbar$-term produced by the derivatives. In this way, each of the equations appearing in (\ref{KE-WF1})--(\ref{KE-WF5}) can be represented as a series in $\hbar$. This series contains real and imaginary terms that should separately vanish. In particular, using the real part of Eq.~(\ref{KE-WF1}) in the leading order of $\hbar$ we obtain
\begin{align}
     k_\mu \mathcal{V}_{(0)}^\mu - \sigma_{(0)} \mathcal{F}_{(0)} + \pi_{(0)} \mathcal{P}_{(0)} = 0
     \label{eq:maincond}
\end{align}
or
\begin{align}
     k_\mu \mathcal{V}_{(0)}^\mu(k) = {\cal M}_{(0)}(k).
\end{align}
This equation should be satisfied for any 4-vector $k$. Thus, we may choose $k=(k^0,0)$. In this case, we may check whether our exact results for $\mathcal{V}^\mu(k)$ and ${\cal M}(k)$ (which are independent of $\hbar$ and should be considered as the leading order terms) satisfy the condition given by \EQ{eq:maincond}. The explicit expressions are
\begin{align}
    &k^0 \mathcal{V}^0(k^0,\kv=0) = \sum_{r=1}^2 k^0 \bigg\{\!\left[ \delta \!\left(k^0 \!-\! 
     E_{\frac{\qv}{2} }^{(r)}\right)\! F_{\frac{\qv}{2}}^{(r)} \!-\! \delta \!\left(k^0 \!+\! E_{\frac{\qv}{2}}^{(r)}\right) \Big(  \bar{F}_{\frac{\qv}{2}}^{\,(r)} \!-\! 1\!\Big) \right]\! \frac{C_{r} \!\left(\frac{\qv}{2},\frac{q}{2}\right)}{2} \nonumber \\
    &\hspace{1.6cm} + \!\left[\delta \!\left(k^0 \!-\! E_{\frac{\qv}{2}}^{(r)}\right)\! F_{\frac{\qv}{2}}^{(r)} \!-\! \delta \!\left(k^0 \!+\! E_{\frac{\qv}{2}}^{(r)}\right) \!\Big(\!\bar{F}_{\frac{\qv}{2}}^{\,(r)} \!-\! 1 \!\Big) \right]\! \frac{C_{r-1} \!\left(\frac{\qv}{2}, \frac{q}{2}\right)}{2} \!\!\bigg\} \label{V00}
\end{align}
and
\begin{align}
    \mathcal{M}(k^0,\kv=0) =  \!\sum_{r=1} ^2 \, \frac{M^2}{ E_{0}^{(r)}}  \left[F^{(r)}_{0}\! \delta\left(k^0 \!-\! E_{0}^{(r)}\right) \!+\! \Big(\bar{F}^{(r)}_{0} \!-\! 1\Big) \delta\!\left(k^0 + E_{0}^{(r)}\right)\right] C_r \left(0, \frac{q}{2}\right) .\label{eq:calMexp0}
\end{align}
We observe that as long as $q \neq 0$, \EQ{eq:maincond} is not fulfilled. This observation shows that the standard semiclassical approach to chiral spirals fails.

\subsection{Alternative expansion}
\label{sec:WF-sc-hbar-alt}

The failure to reproduce the exact result for the Wigner function from the semiclassical expansion presented above suggests that a different expansion scheme in $\hbar$ might be applied. A natural alternative is to consider the case $\kappa = |q|/\hbar \sim \mathcal{O}(1)$. However, this has been already analyzed in~\CIT{Florkowski_1996}, where the pseudoscalar and scalar fields of the form $M(x) \sin(\Phi(x))$ and $M(x) \cos(\Phi(x))$, respectively, were used, with $\Phi(x)$ being an $\hbar$-independent phase. The~results of~\CIT{Florkowski_1996} indicate that there are no modifications of the dispersion relations up to the first order in $\hbar$, which is suggested by our exact result; see~\EQn{E_p^r+} with $q$ replaced by $\kappa \,\hbar$. Consequently, we conclude that the second possible counting scheme is also not able to reproduce the exact result. This suggests the need of the alternative semiclassical methods, for example the approach involving the Moyal product. We leave this topic for possible future investigations.

\section{Summary and Conclusions}
\label{sec:C&O}

In this work, we have obtained exact (up to the quantum loop corrections) expressions for the Wigner function describing spin $1/2$ fermions coupled to an external periodic chiral condensate (chiral spiral). Our calculations were based on the exact solutions of the Dirac equation, to which we applied the method of second quantization. The explicit expressions have been given for all the components of the Wigner function in the so-called Clifford algebra representation. As far as we know, only the case of the free Dirac field has been studied before at the same level of accuracy. The standard method to calculate the Wigner function is the semiclassical expansion. We have demonstrated (for two alternative expansion schemes in $\hbar$) that this approach does not reproduce the properties of the exact Wigner function.

We have found that the formula for the axial vector component of the Wigner function can be conveniently used to study spin polarization effects and illustrate connections between the spin density matrix and axial current. In particular, we have found that during an adiabatic change of the periodic potential into a uniform one, the polarization vector is twisted from its original direction.

\section{Acknowledgments}
\label{sec:Ack}

The authors would like to acknowledge useful discussions with Mariusz Sadzikowski. S. B. would like to acknowledge the support of the Faculty of Physics, Astronomy, and Applied Computer Science, Jagiellonian University via Grant No. LM/36/BS. This work was supported in part by the Polish National Science Centre Grant No. 2022/47/B/ST2/01372 (W. F., S. K. K., and V. M.).

\appendix
\section{Anticommutation relation for the field operators\label{sec:ACrelnDeriv}}
The anticommutation relation of the field operators defined in \EQ{psi-decomp} is given as
\begin{align*}
    &\left\{\psi_a(t,\bm{x}),\psi^\dagger_b(t,\bm{y})\right\}\notag\\
    &= \!\!\!\!\!\sum_{r,s=1,2}\! \int\!\! \frac{d^3\bm{p}\, d^3\bm{p}'}{(2\pi \hbar)^{3}} \frac{1}{2\sqrt{E_{\bm{p}}^{(r)}E_{\bm{p}'}^{(s)}}} \!\left[\!u_a^{(r)}(\bm{p},\bm{x}) u_b^{(s)^\dagger}(\bm{p},\bm{y}) \!\!\left\{\!b_r(\pv)\!, b_s^\dagger(\pv')\!\right\}\! e^{\!- \frac{it}{\hbar} \left(E^{(r)}_p - E^{(s)}_{p'}\right) + \frac{i}{\hbar}(\bm{p}\cdot \bm{x}-\bm{p}'\cdot \bm{y})} \right.\nonumber\\
    &\hspace{5cm}\left.+ v_a^{(r)}(\bm{p},\bm{x}) v_{b}^{(s)^\dagger}(\bm{p},\bm{y}) \!\!\left\{\!c_r^\dagger(\pv)\!, c_s(\pv')\!\right\}\! e^{\!\frac{it}{h} \!\left(E^{(r)}_p - E^{(s)}_{p'}\!\right) - \frac{i}{\hbar}(\bm{p}\cdot \bm{x}-\bm{p}'\cdot \bm{y})}\!\right]\\
    &= \sum_{r=1,2} \!\int\!\! \frac{d^3\bm{p}}{(2\pi \hbar)^{3}\, 2E_{\bm{p}}^{(r)}} \!\left[u_a^{(r)}(\bm{p},\bm{x})u_b^{(r)^\dagger }(\bm{p},\bm{y})e^{i\bm{p}\cdot (\bm{x}-\bm{y})/\hbar} + v_a^{(r)}(\bm{p},\bm{x}) v_{b}^{(s)^\dagger} (\bm{p},\bm{y})e^{-i\bm{p}\cdot (\bm{x}-\bm{y})/\hbar} \right]\\
    &=\! \sum_{r=1,2}\int\! \frac{d^3\bm{p}}{(2\pi \hbar)^{3} 2E_{\bm{p}}^{(r)} }   \left[\left(e^{-\frac{i \gamma_5}{2} \frac{\qv\cdot \xv}{\hbar}}\right)_{ac}\chi_{+,c}^{(r)}(\bm{p}) \chi_{+,d}^{(r)^\dagger}(\bm{p})\left(e^{\frac{i \gamma_5}{2} \frac{\qv\cdot \bm{y}}{\hbar}}\right)_{db}e^{i\bm{p}\cdot (\bm{x}-\bm{y})/\hbar} \right.\notag\\
    &\hspace{5cm}\left.+ \left(e^{-\frac{i \gamma_5}{2} \frac{\qv\cdot \xv}{\hbar}}\right)_{ac}\chi_{-,c}^{(r)}(\bm{p}) \chi_{-,d}^{(r)^\dagger}(\bm{p})\left(e^{\frac{i \gamma_5}{2} \frac{\qv\cdot \bm{y}}{\hbar}}\right)_{db}e^{-i\bm{p}\cdot (\bm{x}-\bm{y})/\hbar}\right]\\
    &= \left(e^{-\frac{i \gamma_5}{2} \frac{\qv\cdot \xv}{\hbar}}\right)_{ac} \left(e^{\frac{i \gamma_5}{2} \frac{\qv\cdot \bm{y}}{\hbar}}\right)_{db}\times\notag \\
    &\hspace{3cm}\sum_{r=1,2} \int \frac{d^3\bm{p}}{(2\pi \hbar)^{3} 2E_{\bm{p}}^{(r)} }\Bigl[\chi_{+,c}^{(r)}(\bm{p})\chi_{+,d}^{(r)^\dagger}(\bm{p}) + \chi_{-,c}^{(r)}(-\bm{p})\chi_{-,d}^{(r)^\dagger}(-\bm{p})\Bigr]e^{i\bm{p}\cdot (\bm{x}-\bm{y})/\hbar}\\
    &= \left(e^{-\frac{i \gamma_5}{2} \frac{\qv\cdot \xv}{\hbar}}\right)_{ac} \left(e^{\frac{i \gamma_5}{2} \frac{\qv\cdot \bm{y}}{\hbar}}\right)_{db} \int \frac{d^3\bm{p}}{(2\pi \hbar)^3}\, \delta_{cd}\, e^{i\bm{p}\cdot (\bm{x}-\bm{y})/\hbar}\\
    &= \left(e^{-\frac{i \gamma_5}{2} \frac{\qv\cdot \xv}{\hbar}}\right)_{ac} \left(e^{\frac{i \gamma_5}{2} \frac{\qv\cdot \bm{y}}{\hbar}}\right)_{db} \delta_{cd} \delta^{(3)}(\bm{x}-\bm{y}).
\end{align*}
This relation is effectively similar to $\delta_{cd} \delta^{(3)}(\bm{x}-\bm{y})$, where indices $c$ and $d$ are spinor indices.
%
\section{Twist of the polarization 3-vector}
\label{ssec:q=0}

In the limit $q\to0$, our spinors $\chi_+^{(r)}(\pv)$ and $\chi_-^{(r)}(\pv)$ become linear combinations of the standard solutions of the free Dirac equation, so we can write~\footnote{Note that we do not list $q$ as the second argument of spinors since its value is external and fixed. In this section $q=0$.}
\begin{align}
\chi_+^{(1)}(\bm{p}) = U^{11} \, u^{(1)}_d(\bm{p}) + U^{12} \, u^{(2)}_d(\bm{p}), \\
\chi_+^{(2)}(\bm{p}) = U^{21} \, u^{(1)}_d(\bm{p}) + U^{22} \, u^{(2)}_d(\bm{p}), 
\end{align}
and, similarly, for $\chi_-^{(r)}(\bm{p})$
\begin{align}
\chi_-^{(1)}(\bm{p}) = V^{11} \, v^{(1)}_d(\bm{p}) + V^{12} \, v^{(2)}_d(\bm{p}) \\
\chi_-^{(2)}(\bm{p}) = V^{21} \, v^{(1)}_d(\bm{p}) + V^{22} \, v^{(2)}_d(\bm{p}). 
\end{align}
The coefficients $U^{rs}$ and $V^{rs}$ can be arranged into $2 \times 2$  unitary matrices $U$ and $V$. For the free Dirac solutions, we use the forms
\begin{align}
    u^{(r)}_d=\sqrt{E_p+M}\begin{pmatrix}
        \varphi^{(r)}\\
        \frac{\bm{\tau}\cdot\bm{p}}{E_p+M} \,\, \varphi^{(r)}
    \end{pmatrix},\qquad v^{(r)}_d=\sqrt{E_p+M}\begin{pmatrix}
        \frac{\bm{\tau}\cdot\bm{p}}{E_p+M}\eta^{(r)}\\
        \eta^{(r)}
    \end{pmatrix},
\end{align}
where
\begin{align}
    \varphi^{(1)}=\begin{pmatrix}
        1\\
        0
    \end{pmatrix},\qquad \varphi^{(2)}=\begin{pmatrix}
        0\\
        1
    \end{pmatrix},\qquad \eta^{(1)} =\begin{pmatrix}
        0\\
        1
    \end{pmatrix},\qquad \eta^{(2)} =-\begin{pmatrix}
        1\\
        0
    \end{pmatrix}.
\end{align}
In what follows, we restrict our discussion to particles only, as the arguments for antiparticles are analogous. 

The axial current in the limit $q \to 0$ includes the matrix element
\begin{equation}
X_{rs} \, {\bar \chi}_+^{(r)}(\bm{p})  \gamma^\mu \gamma_5 \chi_+^{(s)}(\bm{p}), 
\label{eq:me1}
\end{equation}
where $X_{sr}$ is the third Pauli matrix, $X_{sr} = \tau^3_{sr} = X_{rs}$. This is a consequence of our previous assumption that the original spin density matrix is diagonal. Using the matrix notation, we rewrite \EQn{eq:me1} as
\begin{eqnarray}
Mw^\mu_T &\equiv& \sum_{p,q=1}^2 \left( U^\dagger \tau^3 U \right)^{p q} \, \,
{\bar u}_d^{(p)}(\bm{p}) \gamma^\mu \gamma_5 u_d^{(q)}(\bm{p})
= \sum_{p,q=1}^2 W_*^{pq} \, \, {\bar u}_d^{(p)}(\bm{p}) \gamma^\mu \gamma_5 u_d^{(q)}(\bm{p}) \nonumber \\
&=& \sum_{p,q=1}^2
\left( W_*^T \right)^{\,qp} \, \, {\bar u}_d^{(p)}(\bm{p}) \gamma^\mu \gamma_5 u_d^{(q)}(\bm{p}) = \tr \left[ W_*^T
\, \, {\bar u}_d(\bm{p}) \gamma^\mu \gamma_5 u_d(\bm{p})
\right],
\label{eq:me2}
\end{eqnarray}
where trace is taken over spin indices. The symbol $W_*^T$ denotes the transposed matrix $U^\dagger \tau^3 U$, which can be expanded in terms of the Pauli matrices as
\begin{equation}
W_*^T = \sum_{i=1}^3 w^i_{*T} \tau^i ,
\end{equation}
with the coefficients $w^i_{*T}$ defined by the traces
\begin{equation}
 w^i_{*T} = \frac{1}{2} \tr [W_*^T \tau^i].
\end{equation}
The explicit calculation leads to 
\begin{equation}
\wv_{*T} = - \frac{1}{E_p^{\parallel} (E_\pv+M)}\left(p^1 p^3, \,p^2 p^3, \,E_p^{\parallel 2} + E_\pv M  \vphantom{ \frac{1}{2}} \right),
\end{equation}
which gives the normalization $\wv_{*T} \cdot \wv_{*T} = 1$ and $\pv \cdot \wv_{*T} = -E_\pv\, p^3/E_p^{\parallel}$. Finally, the trace in the last line of \EQn{eq:me2} reads (see the formulas elaborated in the Appendix of~\CIT{Florkowski:2019gio})
\begin{equation}
M w^\mu_T = -\frac{4}{E_\pv^\parallel} \left(E_\pv p^3, \, p^1 p^3, \, p^2 p^3, \,(E_\pv^\parallel)^2 \right).
\end{equation}
This 4-vector enters the definition of  the axial current in the limit $q \to 0$ and, due to the orthogonality property $p_\mu w^\mu_T = 0$, its form guarantees that the axial current fulfills the same condition, namely $p_\mu {\cal A}^\mu(\pv) = 0$. Furthermore, it is interesting to observe that $w^\mu_T$ can be obtained by the boost of the 4-vector $w^\mu_{*T} = (0, \wv_{*T})$ from the particle rest frame to the system's CMS frame,
\begin{equation}
w^\mu_T = 4 \, L^\mu_{\,\,\,\nu} \, w^\mu_{*T},
\end{equation}
where $L^\mu_{\,\,\,\nu}$ is the canonical boost defined by the expression~\CITn{Florkowski:2017dyn}
\begin{equation}
L^\mu_{\,\,\,\nu} = 
\begin{bmatrix}
\frac{E_\pv}{M} & \frac{p^1}{M} & \frac{p^2}{M} & \frac{p^3}{M}\\
\frac{p^1}{M} & 1+\alpha_p p^1 p^1 & \alpha_p p^1 p^2 & \alpha_p p^1 p^3\\
\frac{p^2}{M} & \alpha_p p^2 p^1& 1+ \alpha_p p^2 p^2 & \alpha_p p^2 p^3\\
\frac{p^3}{M} & \alpha_p p^3 p^1& \alpha_p p^3 p^2 & 1+ \alpha_p p^3 p^3\\
\end{bmatrix},
\end{equation}
with $\alpha_p = 1/(M (E_\pv+ M))$.

\bibliography{references}
\end{document}